\documentclass{aa}
\usepackage[varg]{txfonts}
\usepackage{graphicx}
\usepackage{natbib}
\usepackage{rotating}
\usepackage{subfigure}
\bibpunct{(}{)}{;}{a}{}{,} 

\newcommand{\grad}{^{\circ}}

\begin{document}
\title{Polarized synchrotron radiation from the Andromeda Galaxy M31 and background sources at 350~MHz}
\author{R.~Gie\ss\"ubel\inst{1}
\and G.~Heald\inst{2}\inst{3}
\and R.~Beck\inst{1}
\and T.G.~Arshakian\inst{4}\inst{5}
}

\institute{
Max-Planck-Institut f\"ur Radioastronomie, Auf dem H\"ugel 69, 53121 Bonn, Germany\label{1}
\and ASTRON, PO Box 2, 7990 AA Dwingeloo, The Netherlands\label{2}
\and Kapteyn Astronomical Institute, University of Groningen, PO Box 800, 9700 AV Groningen, The Netherlands\label{3}
\and I. Physikalisches Institut, Universit\"at zu K\"oln, Z\"ulpicher Straße 77, 50937 K\"oln, Germany\label{4}
\and Byurakan Astrophysical Observatory, Byurakan  378433, Armenia and Isaac Newton Institute of Chile, Armenian Branch\label{5}
}

\abstract
{Low-frequency radio continuum observations are best suited to search for radio halos of inclined galaxies. Polarization measurements at low frequencies allow the detection of small Faraday rotation measures caused by regular magnetic fields in galaxies and in the foreground of the Milky Way.}{The detection of low-frequency polarized emission from a spiral galaxy such as M31 allows us to assess the degree of Faraday depolarization, which can be  compared with models of the magnetized interstellar medium.} {The nearby spiral galaxy M31 was observed in two overlapping pointings with the Westerbork Synthesis Radio Telescope (WSRT), resulting in about 4\arcmin\ resolution in total intensity and linearly polarized emission. The frequency range 310--376~MHz was covered by 1024 channels, which allowed the application of rotation measure (RM) synthesis on the polarization data. We derived a data cube in Faraday depth and compared two symmetric ranges of negative and positive Faraday depths. This new method avoids the range of high instrumental polarization and allows the detection of very low degrees of polarization.} {For the first time, diffuse polarized emission from a nearby galaxy is detected below 1~GHz. The degree of polarization is only $0.21\pm0.05$\%, consistent with the extrapolation of internal depolarization from data at higher radio frequencies. A catalogue of 33 polarized sources and their Faraday rotation in the M31 field is presented. Their average depolarization is $DP(90,20)=0.14\pm0.02$, which is seven times more strongly depolarized than at 1.4~GHz. We argue that this strong depolarization originates within the sources, for instance in their radio lobes, or in intervening galaxies on the line of sight. On the other hand, the Faraday rotation of the sources is mostly produced in the foreground of the Milky Way and varies significantly across the $\sim9$ square degrees of the M31 field.}{As expected, polarized emission from M31 and extragalactic background sources is much weaker at low frequencies than in the GHz range. Future observations with LOFAR, with high sensitivity and high angular resolution to reduce depolarization, may reveal diffuse polarization from the outer disks and halos of galaxies.}


\date{Received 25 April 2013 / Accepted 8 September 2013}
\maketitle

\section{Introduction}

The Andromeda Galaxy (M31) at a distance of $\sim750$~kpc\footnote{For example, $752\pm27$~kpc from the luminosity of the Cepheids \citep{2012ApJ...745..156R} or $744\pm33$~kpc using eclipsing binaries \citep{2010A&A...509A..70V}.} was one of the first external spiral galaxies from which polarized radio emission was detected \citep{beck78}. Most of the radio continuum emission is synchrotron emission, originating from cosmic-ray electrons that spiral around the interstellar magnetic field lines. An early analysis of the polarized emission showed that the turbulent and ordered components of the magnetic field are concentrated in a ring-like structure at about 10~kpc radius \citep{beck82}. This ring is a superposition of several spiral arms with small pitch angles, seen under the inclination of $75\grad$ \citep{chemin}. Faraday rotation measures (RM) showed that the ordered field of M31 is coherent, meaning that it preserves its direction around $360\grad$ in azimuth and across several kiloparsecs in radius \citep{beck82,berkhuijsen03}. This large-scale field can be traced out to about 20~kpc distance from the centre of M31, using polarized background sources in the M31 field as an RM grid \citep{han}.

By now we know that spiral (and even some irregular) galaxies exhibit ordered magnetic fields \citep{beck05} with average field strengths of $5\pm3$~$\mu$G, while the average random field is typically three times stronger (assuming energy equipartition between magnetic fields and cosmic rays) \citep{fletcher10}. In M31 the strengths of the ordered and random fields are about equal ($5\pm1$~$\mu$G) \citep{fletcher04}, which is unique among the galaxies observed so far.

These ordered fields on scales of the entire galaxy can be best explained by dynamo theory \citep{beck96}. It requires weak magnetic seed fields and an interplay of turbulence and shear to generate and maintain such a field. The turbulence can be provided by supernova explosions, while shear is a consequence of the differential rotation of the galactic disk. The exceptionally well ordered large-scale magnetic field in M31 can be well described by an axisymmetric spiral pattern, the basic dynamo mode $m=0$, disturbed by a weaker $m=2$ mode \citep{fletcher04}. The magnetic field in M31 is the prototypical case of a dynamo-generated field.

To our knowledge, there has been no detection of diffuse polarization from spiral galaxies below 1~GHz. All observations of polarization from nearby galaxies were restricted to the GHz range so far. Lower frequencies are advantageous in several aspects: (1) synchrotron emission is stronger, especially from steep-spectrum radio halos, and (2) Faraday rotation is stronger, allowing the measurement of small RM even with low signal-to-noise ratios. Multichannel polarimetry allows the application of RM synthesis \citep{brentjens05}, which generates Faraday spectra for each map pixel. The achieved resolution in Faraday spectra increases with coverage in $\lambda^2$ space ($\Delta \lambda^2$) and hence is higher at low frequencies. On the other hand, Faraday depolarization increases towards lower frequencies at a rate depending on the depolarization mechanism \citep{burn,sokoloff98,tribble}. As a result, polarized emission decreases below some characteristic frequency that depends on the properties of the Faraday-rotating medium \citep{arshakian11}. Depolarization in M31 is relatively low because of its weak turbulent magnetic field (Sect.~\ref{M31pol}). This makes M31 an excellent candidate for low-frequency polarization studies.

The properties of nearby galaxies as observed in radio continuum below 250~MHz will soon be explored with the \emph{LOw Frequency ARray} \citep[LOFAR,][]{LOFAR}, to search for extended synchrotron emission far away from the galactic disks and in the galactic halos. The observations at 350~MHz presented in this paper are a crucial observational link between observations at gigahertz frequencies and the upcoming observations with LOFAR. To this point, the polarization properties of galactic disks and the feasibility to use polarized point sources as a background grid to explore magnetic fields in the foreground are untested at low frequencies.

\section{Observations}

M31 was observed on four different days using the WSRT in December 2008. Two pointings centred at \textsc{Ra}=00h45m00.0s, \textsc{Dec}=41d49m59.9s and \textsc{Ra}=00h41m00.0s, \textsc{Dec}=40d46m00.1s were required to cover the entire galaxy. The maxi-short configuration was used, which is optimized for imaging performance. The shortest baselines in this configuration are 36~m, 54~m, 72~m, and 90~m. Each of the two pointings was observed for $2 \times 12$ hours.

For each 12h pointing four calibrators were observed: 3C295 for flux calibration and 3C303 for polarization calibration at the beginning and 3C147 for flux calibration and DA240 as polarization calibrator at the end.

During the observations the correlator produced 128 channels with a channel width of 78.125~kHz for each of the eight frequency bands in all four cross correlations (XX, XY, YX, YY) with 60~s integration time. The bands (also denoted IFs for intermediate frequencies) were centred on 315.0~MHz, 323.75~MHz, 332.5~MHz, 341.25~MHz, 350.0~MHz, 358.75~MHz, 367.5~MHz, and 376.25~MHz.

After flagging of channels corrupted by radio frequency interference (RFI, see below) the resulting mean frequency of this observation is 343.4~MHz, which corresponds to $\lambda$87.3~cm, called 90~cm throughout this paper.

\section{Data reduction}

Calibration was made in \texttt{CASA 3.3.0} \emph{(Common Astronomy Software Applications)}\footnote{http://casa.nrao.edu/index.shtml} after correcting for the system temperature in \texttt{AIPS} \emph{(Astronomical Image Processing System)}\footnote{http://www.aips.nrao.edu/}. For low-frequency observations with the WSRT there are some limitations of the standard \texttt{CASA} tasks, but since the software is based on the script language \texttt{python}, the user has total control over the data and can also run tasks in batch mode. This is important, since calibration and imaging has to be performed for each channel separately. Due to the large number of channels, this has to be automated.

\subsection{Flagging}\label{wsrt_flag}

With this amount of data, manual flagging of every single channel is no longer possible. The software package \texttt{rficonsole} by \cite{offringa} was used, which was specifically developed for low-frequency data of LOFAR and the WSRT. The algorithm is described in \cite{offringa}. It features a general user interface, with which one can inspect the data manually. Here one develops a so-called strategy (essentially a parameter file used by \texttt{rficonsole}), by randomly checking for single baselines how well the algorithm detects any RFI.

Since the bandpass response across a spectral window is filtered and drops smoothly to zero at the edges to reduce aliasing effects, a bandpass calibration is performed beforehand to facilitate the operation of the algorithm. \texttt{Rficonsole} is applied to the bandpass-corrected data in the \texttt{CORRECTED\_DATA} column, but the flags are stored in a separate table and are also applied to the \texttt{DATA} column, which holds the raw data and is used for the following steps. The final bandpass calibration is made after the flagging. The first 3 and last 17 channels of all IFs are unusable due to the anti-aliasing filter and are flagged as well. Since the bands overlap, one does in general not lose any frequency channels.

However, here two of the bands, IF2 and IF3 (358.75~MHz and 350.0~MHz) were unusable due to RFI and had to be removed entirely for all four days.

Because the maxi-short configuration was used for the observation, antennas RT9 and RTA were subject to shadowing\footnote{see the WSRT shadowing calculator at\\ http://www.astron.nl/\textasciitilde{}heald/tools/wsrtshadow.php}. RT9 was manually flagged for hour angles $\leqq -5$h, RTA for hour angles $\geqq +5$h.

\subsection{Calibration}\label{wsrtcal}

The system temperature ($T_{sys}$) calibration had to be performed in \texttt{AIPS}, since \texttt{CASA 3.3.0} was unable to read the $T_{sys}$ information table provided from the telescope. However, for the calibration in \texttt{AIPS}, the polarization products have to be transformed from linear polarization (XX, XY, YX, YY) to circular polarization (RR, LL, RL, LR). Detailed instructions are given in the CookBook for WSRT data reduction using classic \texttt{AIPS} by R.~Braun\footnote{http://astron.nl/radio-observatory/astronomers/analysis-wsrt-data/analysis-wsrt-dzb-data-classic-aips/analysis-wsrt-d}.

The observation of 3C295 and 3C303 failed on the second day. For consistency only 3C147 and DA240 were used for calibration. Where possible, the gain solutions were applied to 3C295 and 3C303 to check their validity.

The flux of 3C147 was derived using the analytic function given in the VLA Calibrator Manual\footnote{http://www.vla.nrao.edu/astro/calib/manual/baars.html} \citep{vlacal},
\begin{equation}
\log(S_\nu)=A+B \log(\nu) + C \log^2(\nu) + D \log^3(\nu),
\end{equation}
with $A=1.44856,  B=-0.67252,  C=-0.21124,  D=+0.04077$. At the time of writing, more precise models for the six most common calibrators at low frequencies were published by \cite{scaife12}. For 3C147 both models agree well within the uncertainties, but the spectrum is probably slightly steeper in our frequency range.
The deviation is lower than 3\%, therefore we did not need to repeat the entire data reduction.

For DA240 the values published in \cite{brentjens08} were assumed: $RM=+3.33\pm0.14$~rad~m$^{-2}$ and a polarization angle at $\lambda^2=0$ of $pa=122\grad\pm3\grad$.

At these wavelengths 3C303 is expected to be 5\% polarized with an RM of +15~rad~m$^{-2}$ (Ger de Bruyn, private communication). This value differs slightly from (but still agrees with) the published value of $RM=+18 \pm 2$~rad~m$^{-2}$ at GHz frequencies \citep{simard}.

The results of calibrating 3C303 confirm that the polarization calibration remains constant over the the 12~h of each observation, which means that the ionosphere was reasonably stable during this time span. In December 2008, solar activity was near a minimum, ionospheric Faraday rotation is thus expected to be only a few rad~m$^{-2}$ and stable, which means that any effects can be handled by the calibration.

The calibration has to be made for each channel individually, solving for all elements in the Jones matrices \citep{hamaker}, to circumvent problems caused by the so-called 17~MHz ripple. This is a variation of the gains with frequency (at a period of 17~MHz), caused by a standing wave between the dish and the primary focus of the antennas.
It sometimes results in strong frequency-dependent variations in the spectra of off-axis sources across the primary beam and polarization leakages \citep{poppingbraun,brentjens08}.

The properties of the calibrators (namely the expected values for the four Stokes parameters) were calculated for each channel independently and were written into the \texttt{MODEL\_DATA} column of the measurement set. Afterwards the \texttt{CASA} tasks \texttt{gaincal}, \texttt{polcal}, and \texttt{applycal} were used to calculate and apply the gains to the other calibrators and the M31 fields. This was done separately for each channel and each of the four observations.

\subsection{Selfcal, peeling, and imaging}\label{selfandpeel}

Several rounds of self-calibration (selfcal) were performed before final imaging of the visibility (uv) data of M31. A single selfcal step consists of imaging and cleaning the uv data to obtain a clean model of the field and using that clean model for calculatin and applying gain solutions to the field. But first Cassiopeia A and Cygnus A had to be removed from the data, since during imaging side-lobes from both sources were visible within the M31 field. The two sources are far away from the pointing centre, but they are the brightest sources in the sky at these frequencies.

Sources can be subtracted from the uv data using the so-called peeling method. This is a general method for removing off-axis sources from the field, for which good gain solutions cannot be derived with normal selfcal. It requires the following steps:
\begin{enumerate}
\itemsep0pt
\item{Obtain a good clean model of the observed field. Usually a selfcal step (excluding the source(s) to be subtracted) is performed to be able to produce a better clean-model.}
\item{Subtract that model from the uv data. This step is performed so that no side-lobes from the field interfere with the source that is to be peeled. If a selfcal step was made, the model has to be subtracted from the \texttt{CORRECTED\_DATA} column. Since the gain-solutions from the selfcal step did not include the source that is being peeled, they will be entirely wrong for this source. Thus, after subtracting the calibration with the field sources has to be reversed by inverting the gain table and applying the inverted gains.}
\item{Use the field-subtracted uv data to obtain a good clean model of the source that is to be peeled. Usually this involves setting the phase centre to the position of the source, and again performing selfcal on the source in question.}
\item{Subtract the new model from the original uv data. If selfcal was used, first apply the gain solutions, then subtract and reverse the calibration by again applying the inverted gains.}
\end{enumerate}
For better results, this can be repeated in an iterative process, since the model for the field will improve after the disturbing influences of the peeled source are reduced, leading to a better subtraction and thus a better model for the source to be peeled, and so on.

Here it was sufficient to run selfcal on the field using only the brightest sources, subtract that model from the field, apply the inverted gains from the selfcal and then subtract Cassiopeia~A and afterwards subtract Cygnus~A without any selfcal steps. (Both sources are too far away from the original phase-centre to achieve any good solutions.)

After peeling, three selfcal runs were performed on each M31 field (again separately for each channel and each of the four days). For the initial model, the uv range was restricted to baselines $>0.1$~k$\lambda$ to exclude the extended emission and start with a simple model. Only the brightest sources and uniform weighting were used. The resulting image was cleaned to three times the noise-level measured in the Stokes V image for each channel individually. In the second run the uv range was still restricted, and more sources as well as the bright centre of M31 were included. The threshold for cleaning was reduced to twice the noise. In the final step the uv range was no longer restricted and the entire M31 was included in the model and cleaned down to the noise-level.

This procedure was tested for different channels across the entire bandwidth and then automated with \texttt{python} to run without user-interaction for each channel separately.

\subsection{Final imaging}\label{finalimaging}

For the final images the uv range was restricted from 34.5~$\lambda$ to 2801.0~$\lambda$. This is the maximum uv range that all channels have in common, and ensures that for all frequencies the spatial resolution is the same. The images for all four Stokes parameters were cleaned automatically using Briggs-weighting and a uv taper and afterwards were slightly smoothed to a final resolution of $230''\times290''$. The Stokes Q and U images are then used for RM synthesis (Sect.~\ref{RM}).

For the total power map, the channels were once again inspected using the images from the automatic imaging run, and obviously bad channels were excluded. Then the uv data was concatenated per field and per IF, keeping the individual channels, and imaged manually, resulting in one image per field and per IF. As a last imaging step the two fields had to be mosaicked and primary-beam corrected.

\begin{figure*}[htb!]
\begin{center}
\includegraphics[trim=0 0 0 0,clip,scale=0.6,angle=-90]{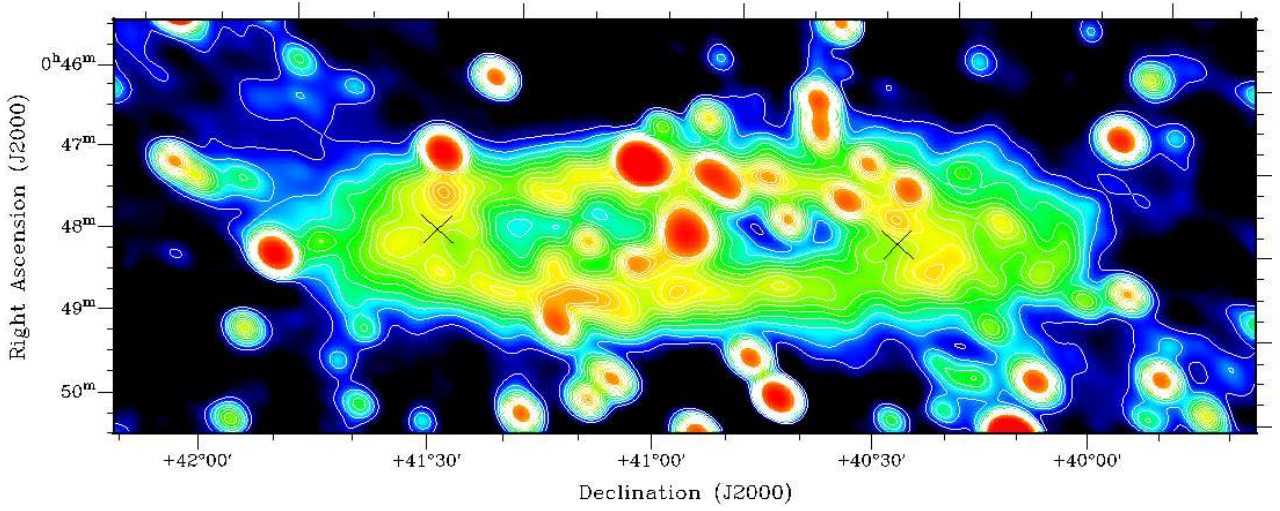}
\caption[Total power image of M31 at 90~cm.]{Final total power image of M31 at 90~cm. Contours are from 4 to 128~mJy/beam in steps of 10~mJy/beam. HPBW: $230''\times290''$; rms=1.4~mJy/beam. The black crosses mark the pointing centres of the two fields. Note: The tick marks may be misleading since the map is rotated with respect to the celestial coordinates, see the coordinate grid in Fig.~\ref{M31_SI} or Fig.\ref{rm_map}!}
\label{M31_wsrt}
\end{center}
\end{figure*}

\begin{figure*}[hbt!]
\begin{center}
\includegraphics[scale=0.6,angle=-90]{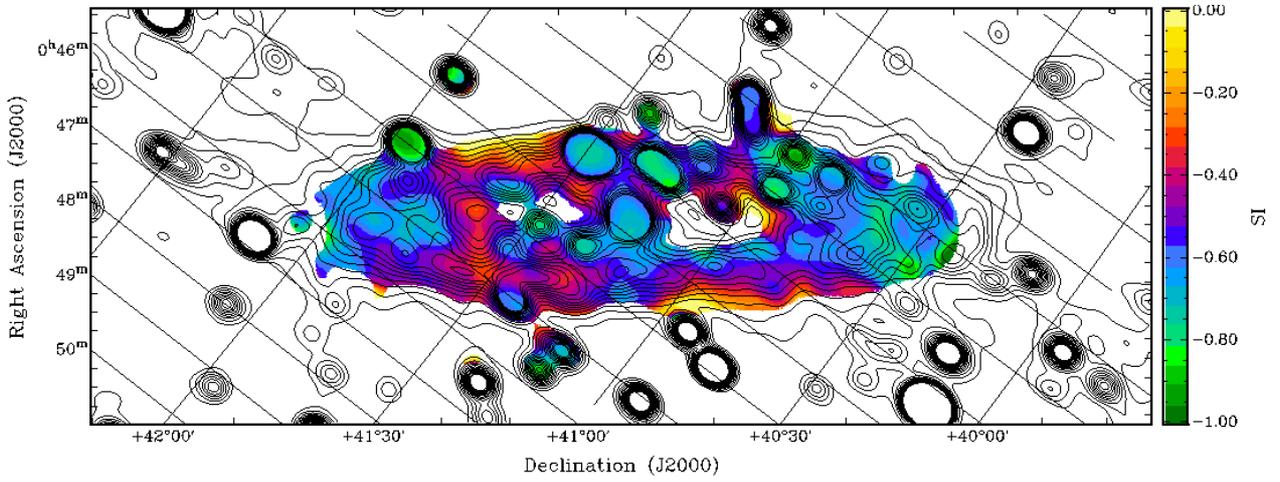}
\caption{Spectral index (SI) map between 90~cm and 20~cm overlaid with the same contours as Fig.~\ref{M31_wsrt}. The SI map was calculated for intensities $>5\sigma$ in Stokes I.}
\label{M31_SI}
\end{center}
\end{figure*}

\subsubsection{Primary-beam correction and mosaicking}

According to the ASTRON webpage, the primary beam response can be described by the function\footnote{http://astron.nl/radio-observatory/astronomers/wsrt-guide-observations/wsrt-guide-observations}
\begin{equation}\label{pbeam}
A(r)=cos^6(c \nu r),
\end{equation}
where $r$ is the distance from the pointing centre in degrees and $\nu$ the observing frequency in GHz. The automatic primary beam correction in \texttt{CASA} uses a different function for the WSRT. The constant $c$ is only constant for GHz frequencies and is said to decline to $c=66$ at 325~MHz and $c=63$ at 4995~MHz. It is related to the light crossing time across the the aperture (i.e. the effective diameter of the dish, \citealt{brentjens08}). A good primary beam correction is crucial in our case, because M31 spans two pointings. Since there is no exact value given and recommendations vary (e.g. $c=64$, Ger de Bruyn, private communication), we manually determined the best value for our observation.

For four arbitrary chosen channels across all bands (IF 0, channel 15; IF 1, channel 83; IF 5, channel 26; and IF 7 channel 57), the primary-beam correction was applied for different values of $c$. Then the difference between the peak fluxes of each of the ten sources in the overlap area of the primary beams between the northern and southern field at the same frequency was calculated and normalized to the flux density in the northern field,
\begin{equation}
f_i(c)=\frac{S_\nu^{i,north}-S_\nu^{i,south}}{S_\nu^{i,north}}.
\end{equation}
Hence, for each constant $c$ there are ten values $f_i(c)$ (per chosen channel) between -1 and +1. For a perfect primary-beam correction, each value would be equal to 0. The error in the mean of these ten values is thus a measure for the quality of the primary beam correction,
\begin{equation}
\sigma(c)=\sqrt{\frac{1}{n(n-1)}\sum_i^n \left( \overline{f(c)} - f_i(c) \right)^2}.
\end{equation}

For $c=65$, all the systematic errors are at their lowest common point and are also all below 10\%, which is deemed an acceptable level. Using this value for the primary beam correction, the resulting systematic flux error was estimated to be $7-8\%$.

Mosaicking was made in the image domain, using a standard linear mosaicking scheme \citep[see][]{cornwell,sault_mosaic}, assuming equal noise levels for both fields,
\begin{equation}
I_{LM}(l)=\frac{\sum_p A(l-l_p)I_p(l)}{\sum_p A^2(l-l_p)},
\end{equation}
where $A(l-l_p)$ is the primary beam attenuation at distance $l$ from the pointing centre $l_p$ (eq.~\ref{pbeam}) and $I_p$ is the cleaned image of the respective field.

\section{Total emission from M31 at 90~cm}\label{sec:total_em}
The final total power image is shown in Figure~\ref{M31_wsrt}. It is a uniformly weighted average of the single IF images (see Section~\ref{finalimaging}). Note that even at these low frequencies, no excess radio continuum emission (i.e. a radio halo) is detected around M31. This has already been inferred by \cite{graeve81}.

After subtracting all sources $< 0.1$ Jy, we found a total flux density integrated over the radius interval $R=0-17.4$~kpc\footnote{Previous measurements used the radius interval $R=0-16$~kpc based on the old distance estimate of 690~kpc by \cite{vaucouleurs}, which corresponds to $R=0-17.4$~kpc using the current estimates by \cite{2012ApJ...745..156R} and \cite{2010A&A...509A..70V}.} of $10.6\pm0.7$ Jy. The error is estimated from the standard deviation of the total fluxes for M31 for the individual IF images. The value is consistent with the integrated flux densities listed by \cite{berkhuijsen03}. We note that the value may be slightly too low due to missing spacings. After restricting the UV range (see Section~\ref{finalimaging}), our largest detectable structure is $1.7\grad$. This would still enable us to detect a halo along the minor axis.

Figure \ref{M31_SI} shows a spectral index (SI) map between our 90~cm map and the VLA+Effelsberg 20~cm map by \cite{vla20}. The spectral index is very similar to that presented by \cite{berkhuijsen03} between 20~cm and 6~cm. Like at GHz frequencies, we found on average a slightly steeper spectral index towards the southern major axis ($\alpha_{90,20}\approx-0.7$) compared with the northern major axis ($\alpha_{90,20}\approx-0.6$). The north-eastern part of the ring (lower left in figure~\ref{M31_SI}) is dominated by H\textsc{ii} regions and shows a flat spectral index of only $\alpha_{90,20}\approx-0.4$. This is even flatter than the average value found in the SI map by \cite{berkhuijsen03} at GHz frequencies. With a constant thermal fraction and a constant nonthermal spectral index one would expect the spectral index to steepen at low frequencies, so this indicates that thermal absorption is considerable. We note that missing spacings are no problem for the spectral index here, since they only become significant in the faint outermost regions. The overall spectral index, dominated by the bright regions, is therefore not affected.

A more thorough analysis would require a higher resolution map at 90~cm to allow proper subtraction of all point sources and is beyond the scope of this paper.

\section{Polarized emission and RM synthesis}\label{sec:rmsyn}
\label{RM}
The Faraday depth (FD) is proportional to the integral along the line of sight over the cosmic-ray electron density $n_e$ and the strength of the line of sight component of the regular magnetic field $\vec{B}$ \citep{burn},
\begin{equation}
\phi \propto \int_{\text{source}}^{\text{observer}}n_e\: \vec{B}\cdot \text{ d}\vec{l},
\end{equation}
while the classical RM is an observable quantity that describes the difference of polarization angles $\Delta\chi$ observed at two (or more) different wavelengths $\lambda_i$,
\begin{equation}
\text{RM}=\frac{\Delta\chi}{\lambda_1^2-\lambda_2^2}\label{eq:rm}.
\end{equation}
The classical RM is equivalent to the FD $\phi$ only if there is just a background source and a dispersive Faraday screen in the foreground along the line of sight. If there are several emitting and rotating components along the line of sight, the linear relationship between $\Delta\chi$ and $\Delta(\lambda^2)$ in eq.~\ref{eq:rm} does not hold. In addition, there is an ambiguity because the polarization vector could have rotated by $n\pi$ ($n$ a natural number) without being noticed.

\begin{figure}[hbt!]
\centering
\subfigure[Sketch of different components along the line of sight. Some of them are emitting ($Ei$), are Faraday-rotating ($Ri$), or both.]{
    \includegraphics[scale=0.28]{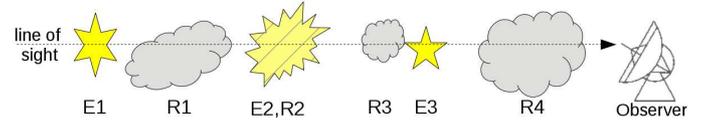}
    \label{LOS}
}
\subfigure[Resulting Faraday spectrum of the components depicted in~\ref{LOS}.]{
    \includegraphics[scale=0.4]{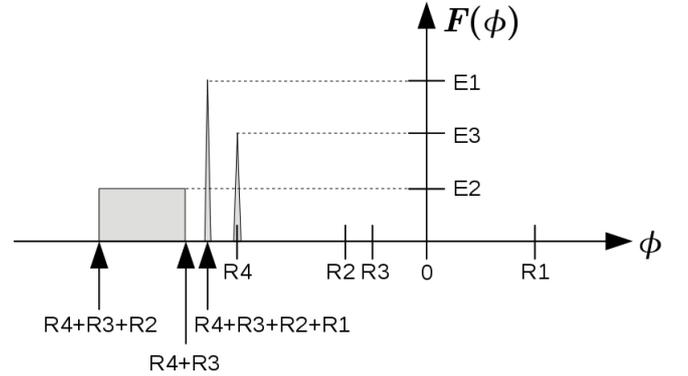}
    \label{LOS-spectrum}
}
\caption{Example for a set of different polarization-emitting and rotating components along the line of sight and the resulting Faraday spectrum. $E1$ and $E3$ are point sources, $E2,R2$ is an extended emitting and rotating region. $R1$, $R3$, and $R4$ are Faraday screens that are not emitting. Depolarization effects are neglected here for simplicity.}
\label{fig:rmsynth}
\end{figure}

RM synthesis, on the other hand, yields a spectrum $F(\phi)$ in FD $\phi$, in which each polarization-emitting component along the line of sight will produce a separate signal. Its position in FD corresponds to the total rotation of the rotating components between the emitting component and the observer along the line of sight (see Fig.~\ref{fig:rmsynth}). For more details on RM synthesis we refer to \cite{brentjens05} and \cite{healdRM}.

\subsection{RM synthesis at 90~cm}

RM synthesis was performed on the automatically cleaned Q and U images after mosaicking, using the software by Michiel Brentjens \citep{brentjens05,brentjens08}. The resulting Faraday cube was then cleaned using the RM~Clean code for \texttt{MIRIAD} \citep{miriad} implemented by one of us (GH) \citep{sings2}, using 1 $\sigma$ as cutoff level and $nmax=1000$. RM~clean converged after $\sim975$ iterations.

The resolution in FD $\phi$ of our observation is given by the measured full width at half maximum of the RM spread function (RMSF) $\varphi=19.12$~rad~m$^{-2}$. The largest detectable structure in Faraday space is $\phi_{max}=\pi/{\lambda_{min}^2}\approx4.95$~rad~m$^{-2}$.

Note that the terms RM clean and RMSF are misleading because only in exceptional cases RM and FD $\phi$ are the same (see above).

\subsection{Polarized sources}
\subsubsection{Catalogue}\label{catalogue}

\begin{figure}[h]
\begin{center}
\includegraphics[trim=0px 0 0 0,clip,scale=0.4,angle=-90]{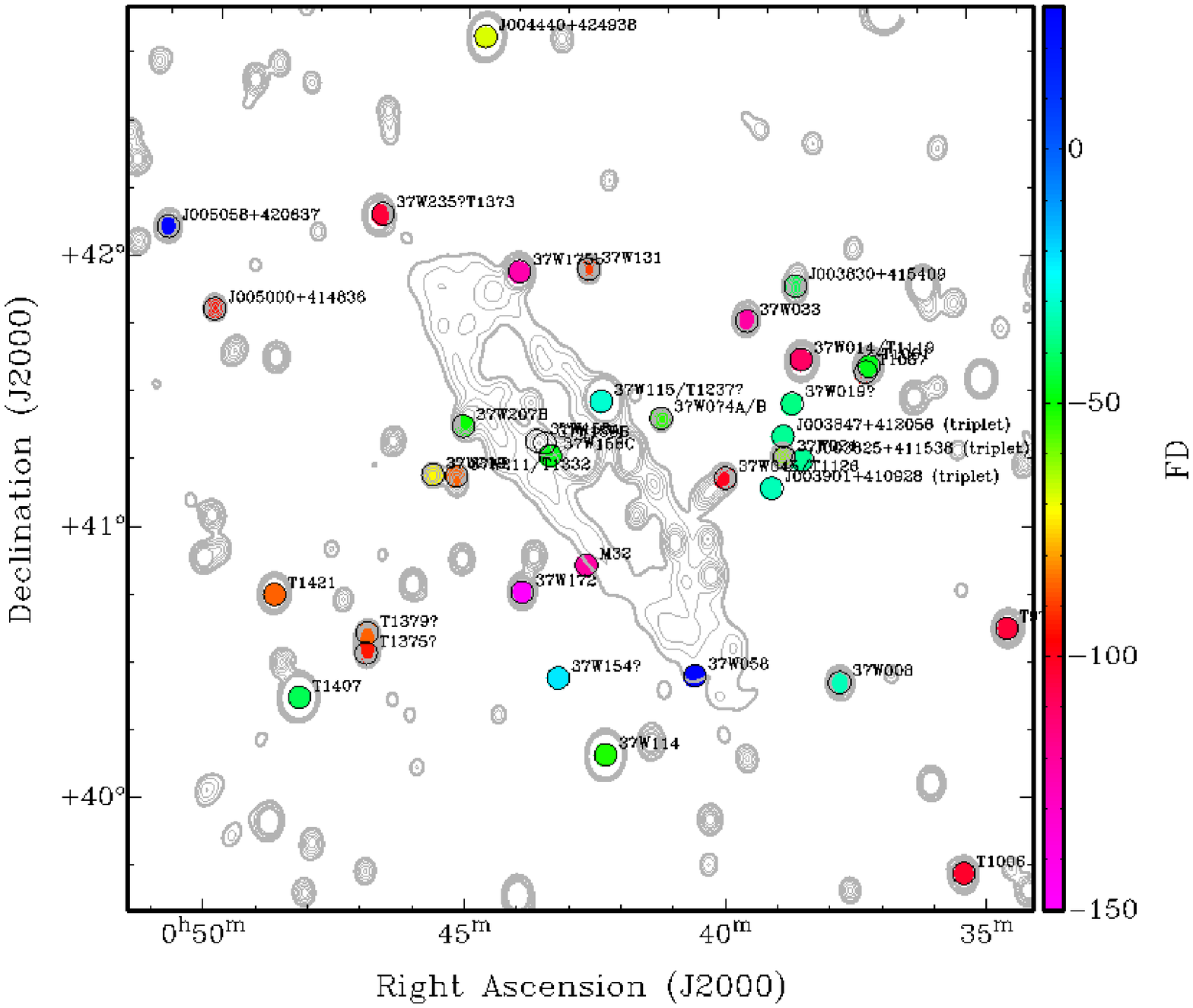}
\caption[Map of the FD sources.]{Position and Faraday depth of the detected sources on contours of the total power map.}
\label{rm_map}
\end{center}
\end{figure}

We compiled a catalogue of all (33) polarized sources detected in the Faraday cube in Table~\ref{RMcat}. The listed positions correspond to the brightest, central pixel of that source in the image plane at the FD it was detected. The polarized intensities (PI) are the peak intensities in the Faraday cube, and each polarization degree ($p$) was calculated with the respective intensity at that position in the Stokes I map. The name of the corresponding source from the \cite{walterbos} catalogue and/or from \cite{taylor} is given if the source is listed. If marked with a `?' the assignment is not entirely certain because the position of the peak and that in the catalogue differ slightly. This can occur because of uncertainties due to the low resolution in the WSRT data, sources consisting of several unresolved components, or in fact an erroneous assignment. For comparison the polarization properties measured at 1.4~GHz (21~cm) by \cite{han} and \cite{taylor} are listed where available. We note that only nine of the 21 sources listed by \cite{han} were detected at 90~cm. The FD derived from RM synthesis at 90~cm is denoted by $\phi$, whereas $RM$ denotes Faraday rotation measures obtained from two frequency bands.

The average value of $\phi$ of all sources in the catalogue is $-67\pm40$~rad~m$^{-2}$ (where $\pm40$~rad~m$^{-2}$ is the standard deviation). The average degree of polarization is 1.36\%. Most values are below 3\%; only five sources are more strongly polarized.

An important selection criterion for a source to be included in the catalogue was a clear recognition as a point-like source in the cube. Plots of the FD along slices through the source positions in \textsc{Ra}/\textsc{Dec} are a good tool for determining the spatial extent of peaks in the Faraday cube. In Appendix~\ref{spectra.app} the Faraday spectra of all detected sources are shown. The polarization data (Figures \ref{rm_spec1} to \ref{rm_spec6}) is available in FITS format at the CDS\footnote{\tt http://cdsweb.u-strasbg.fr/cgi-bin/qcat?J/A+A/}.
Some have several components, but these peaks belong to extended structures in the cube that are unrelated to M31 (Sect.~\ref{M31pol}). The error in FD is estimated by
\begin{equation}
\Delta \phi=\frac{\varphi}{2 S / N},
\end{equation}
where $\varphi$ is the full width at half maximum of the RMSF and $S/N$ the signal-to-noise ratio of the peak. This is a similar error estimation as is used for source detection in imaging (e.g. \citealt{fomalont}). The error in polarized intensity is the noise level of the Faraday cube, which is about 1/${\sqrt{N_{chan}}}$ times lower than the rms in the single Q and U images ($N_{chan}$ is the number of frequency channels). The lowest $S/N>8$, therefore polarization-bias correction can be neglected.

Figure~\ref{rm_map} shows the position of the detected sources on contours of the total power map. The colour corresponds to the measured FD. There are clearly not enough sources to define an RM grid, which could be used to trace M31 and its magnetic field as a foreground screen \citep{han,stepanov}. However, 37W175b in the north, 37W207B, 37W115/T1237? and 37W158C in the middle, and 37W058 in the south seem to follow the RM distribution seen in M31 at GHz frequencies \citep{berkhuijsen03}.

The average RM value for the foreground determined at GHz frequencies is -93~rad~m$^{-2}$ \citep{fletcher04}, consistent with the average $\phi$ of the 90~cm sample. However, towards the west a number of sources seem to increasingly show a $\phi$ of about -50~rad~m$^{-2}$. This may be an indication that the foreground screen is in fact not constant, but has an FD gradient. However, the western part of the cube is heavily affected by foreground emission, as can be seen from the Faraday spectra of the sources (Sect.~\ref{spectra.app}).

\begin{figure}[ht!]
\centering
\subfigure[$\phi=-44$~rad~m$^{-2}$]{
    \includegraphics[trim=29px 0px 0px 0px,clip,scale=0.28,angle=-90]{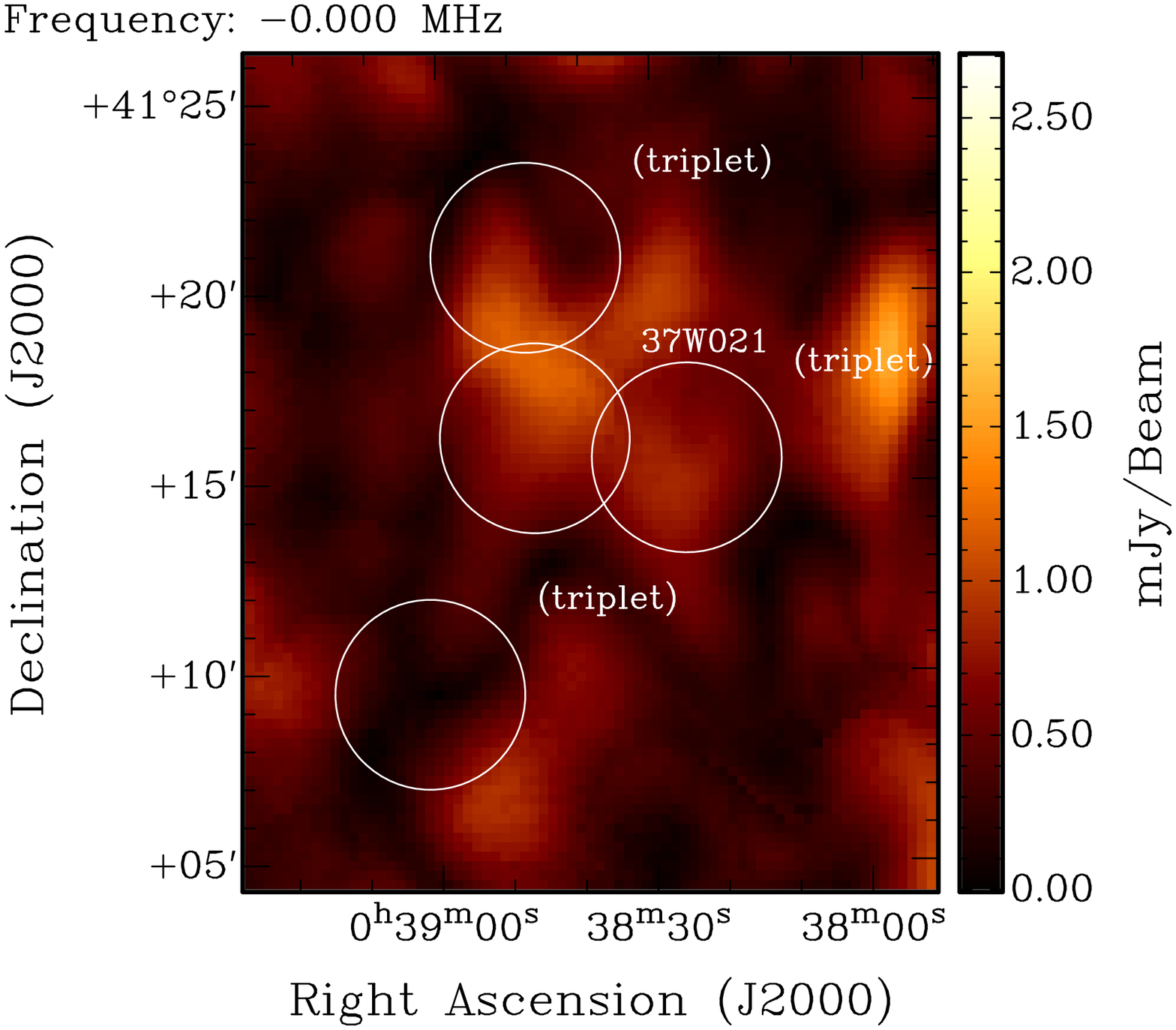}
    \label{triplet1}
}
\subfigure[$\phi=-40$~rad~m$^{-2}$]{
    \includegraphics[trim=29px 0px 0px 0px,clip,scale=0.28,angle=-90]{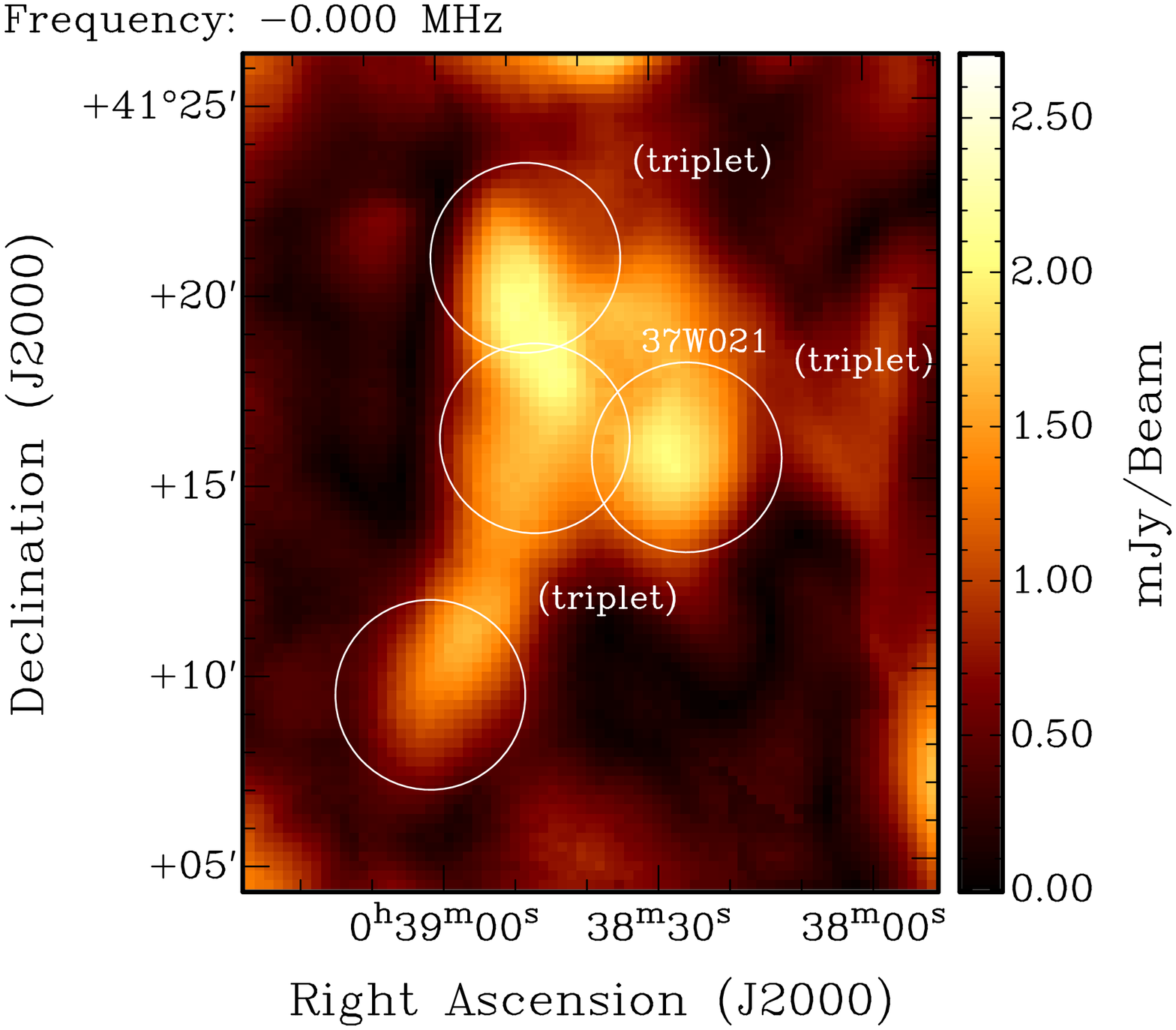}
    \label{triplet2}
}
\subfigure[$\phi=-36$~rad~m$^{-2}$]{
    \includegraphics[trim=29px 0px 0px 0px,clip,scale=0.28,angle=-90]{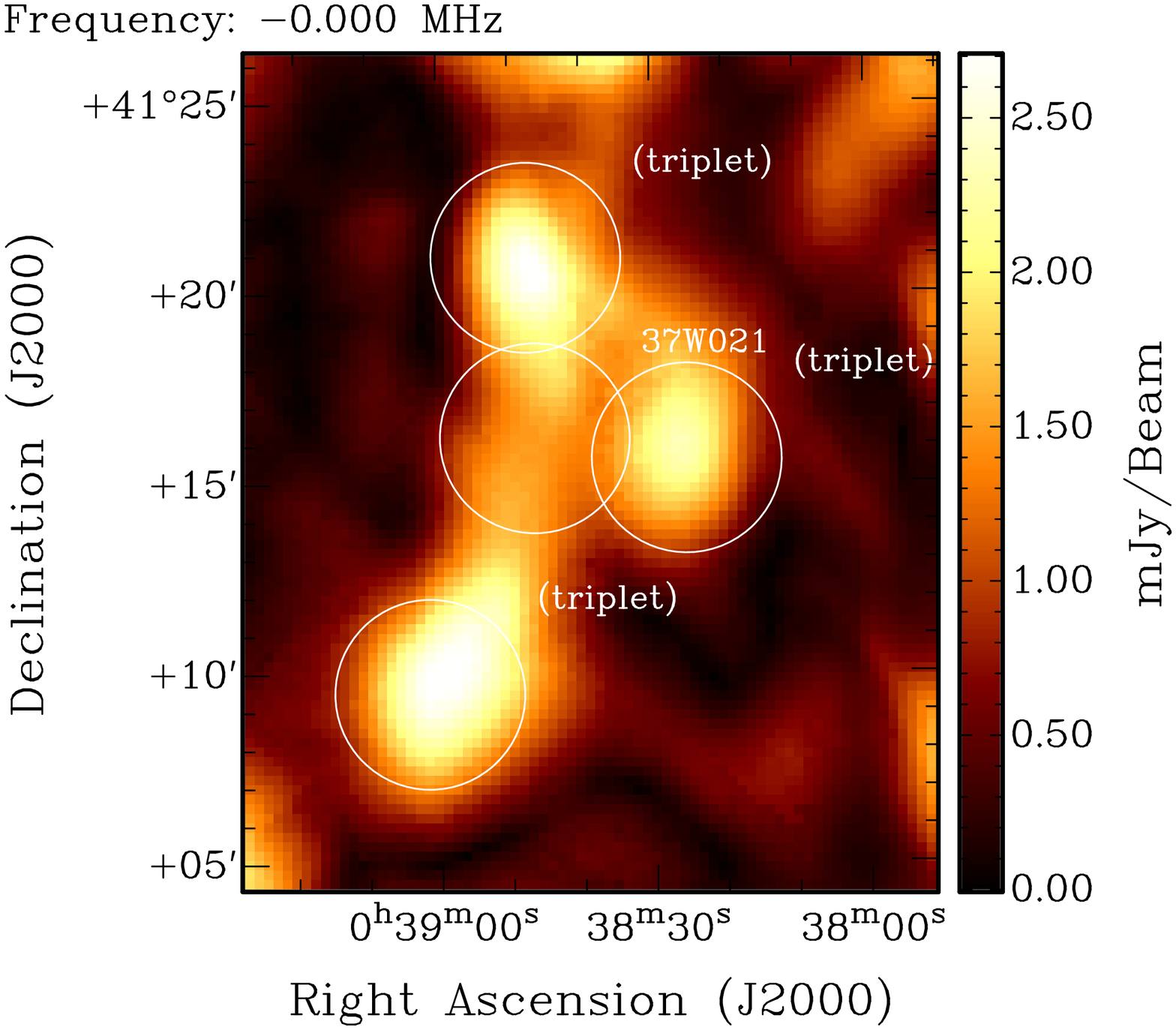}
    \label{triplet3}
}
\subfigure[$\phi=-32$~rad~m$^{-2}$]{
    \includegraphics[trim=29px 0px 0px 0px,clip,scale=0.28,angle=-90]{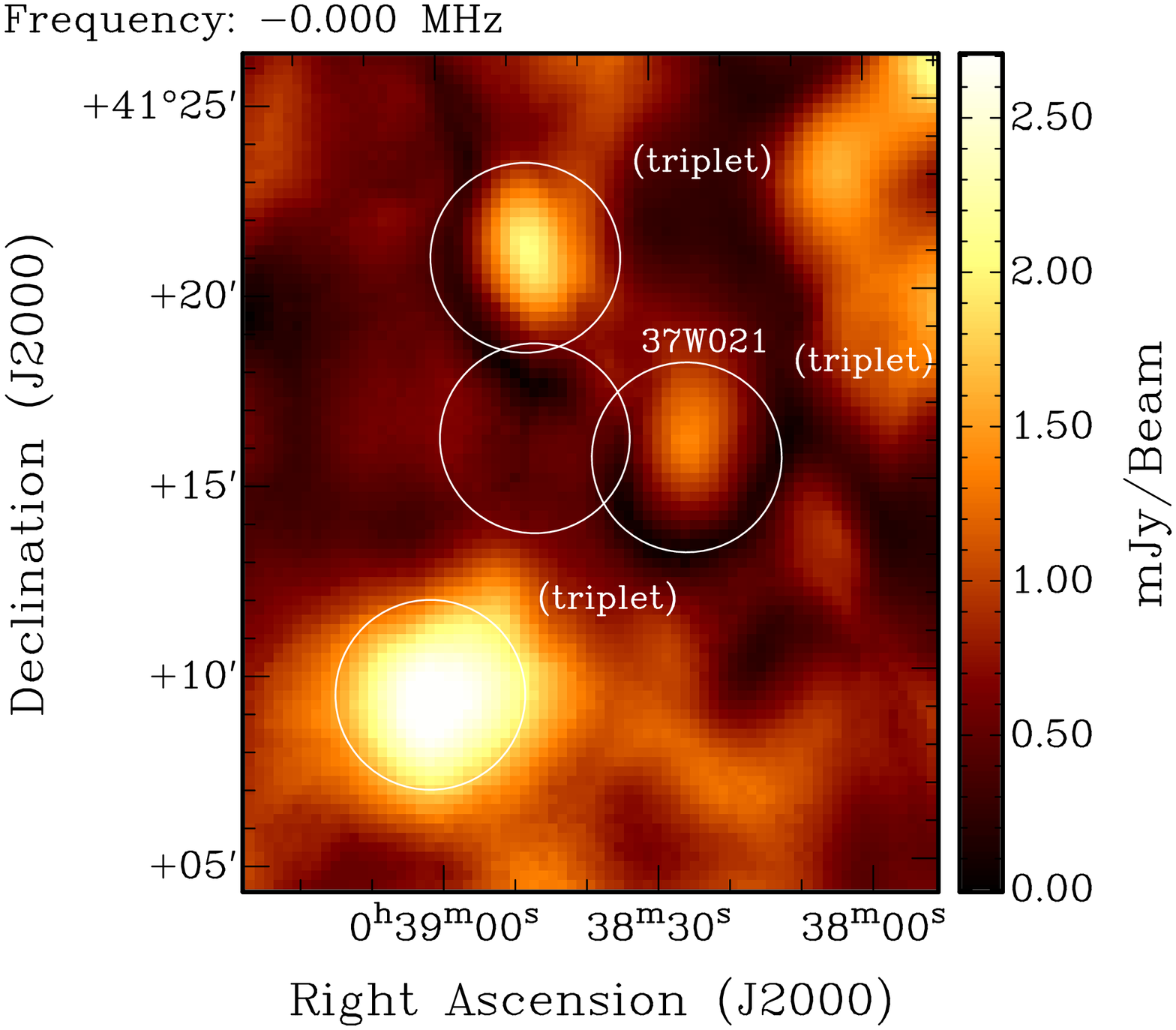}
    \label{triplet4}
}
\caption[The ``triplet'' feature.]{Frames from the Faraday cube of the triplet feature at different FDs.}
\label{triplet}
\end{figure}

Moreover the three sources marked as (triplet) (J003901+410928, J003847+412056 and J003825+411538) are probably not background sources, but are part of a peculiar feature in the cube, possibly polarized emission from the Milky Way. Figure \ref{triplet} shows several frames of the Faraday cube at different FDs of that feature. The feature emerges at the location of 37W021 and extends star-shaped to the denoted positions. It seems unlikely that there is a physical connection to 37W021 because of their relatively large angular separation and the lack of counterparts in total intensity. If these features are truly connected, they are probably part of the Galactic foreground.

There are two ubiquitous features, detected at most pixels in \textsc{Ra}/\textsc{Dec}: a peak at $\phi=0$~rad~m$^{-2}$, which is caused by the frequency-independent instrumental polarization, and a peak around $\phi=-12$~rad~m$^{-2}$ (sometimes accompanied by one or two more peaks within the range -30 to 0~rad~m$^{-2}$). This could be polarized emission from the near side of the foreground medium of our Milky Way. Since this hampers the detection of signals from background sources or M31 in the Faraday cube, the range $\pm30$~rad~m$^{-2}$ is marked grey in all plots. An emitting foreground region is expected to be recognizable as an extended feature in the Faraday spectrum, but is not entirely visible in our observations. With a shortest wavelength of $\lambda_{min}\approx80$~cm, the widest detectable feature in Faraday spectra has an extent $\text{FD}_{max}=\pi/\lambda_{min}^2\approx5$~rad~m$^{-2}$. 
A pair of Faraday components with similar heights would indicate an extended FD structure. The quality of our data is not sufficient to rule out such structures. A large coverage in $\lambda^2$ is needed to recognize components with a range of scales in the Faraday spectrum \citep[e.g.][]{brentjens05,beck12}.

\subsubsection{Comparison with previous works}\label{compar}

\begin{figure}[h]
\begin{center}
\includegraphics[scale=0.45]{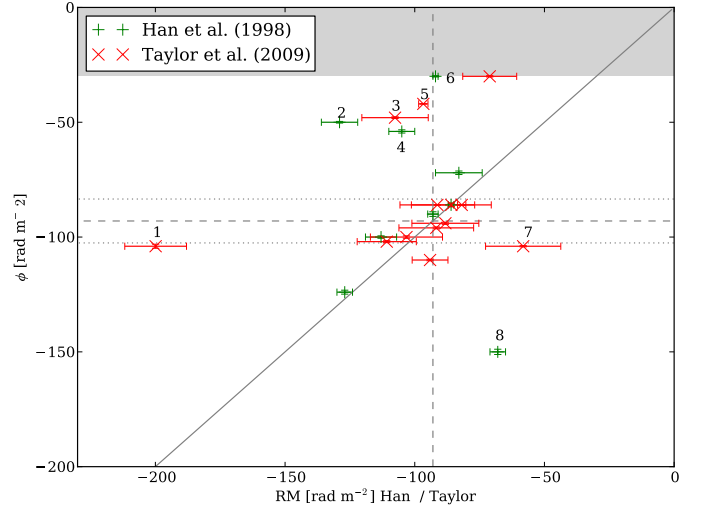}
\caption[RM comparison with literature.]{Measured FD from the Faraday cube at 90~cm versus RM values measured by \cite{han} and \cite{taylor} at 20~cm. The grey area marks the range of FD that probably is affected by foreground emission and instrumental polarization. Dashed lines indicate the expected rotation measure of the Galactic foreground, dotted lines show the range spanned by the FWHM of the RMSF.}
\label{RMcompare}
\end{center}
\end{figure}

Not all sources detected in the Faraday cube have been listed by \cite{han} or \cite{taylor} and vice versa. For the subset of sources appearing in multiple catalogues, the measured rotation measures and polarized intensities can be compared. In Figure~\ref{RMcompare} the FD detected in our catalogue is plotted against the RM values found by \cite{han} (green) and \cite{taylor} (red). The solid line indicates where both values coincide. Although most sources agree well, several sources deviate. Here they are identified with numbers from 1 to 8, which are also given behind the names of the corresponding source in Table~\ref{RMcat}. Sources 1 to 6 are systematically offset from the solid line towards lower absolute FDs, which suggests a systematic effect caused by Faraday depolarization.

The RM values by \cite{han} and \cite{taylor} were determined in the traditional way, using only two frequencies (1.365~GHz and 1.652~GHz by \citealt{han}, 1.365~GHz and 1.435~GHz by \citealt{taylor}). Although the frequency pairs are very close, some of their values could still be affected by the $n\pi$ ambiguity. However, this would yield an offset of $\sim\pm650$~rad~m$^{-2}$, which is too high to explain the offsets seen in Fig.~\ref{RMcompare}. (Moreover \cite{han} and \cite{taylor} do not always in agree (see e.g. source 6 (37W115/T1237? or 37W045/T1126) where the PI values disagree by an order of magnitude.)

Inspection of the Faraday spectra of the sources (Figs.~\ref{rm_spec1}--\ref{rm_spec6}) suggests that most of the spectra of the deviating sources have a more complex structure than those where the $\phi$ value matches the RM from previous works. The simple linear approximation of the variation of polarization angle with $\lambda^2$ is only valid for the simplest case of a background source without rotation and a constant foreground screen without emission (see above). For example, the spectra of 37W074A/B (Figure~\ref{rm_spec3}, bottom right) and T1061 (Fig.~\ref{rm_spec6}, top left) suggest the presence of several possibly extended RM structures along the line of sight, while the spectra of 37W175b (Fig.~\ref{rm_spec2}) or 37W045/T1126 (Fig.~\ref{rm_spec4}), for instance, are much more simple. Multiple components along the line of sight can easily lead to wrong results when measuring the RM between only two frequencies, which probably explains the deviations in Fig.~\ref{RMcompare}.

There may also be systematic errors in the output of RM synthesis \citep[e.g.][]{farnsworth}. However, such ambiguities are within the FWHM of the RMSF. As can be seen in Figure~\ref{RMcompare}, the offset sources are well outside of this range, so they are unlikely to be caused by such systematic uncertainties resulting from RM synthesis of complicated sources.

\subsubsection{Depolarization}

\begin{figure}[h]
\begin{center}
\includegraphics[scale=0.45]{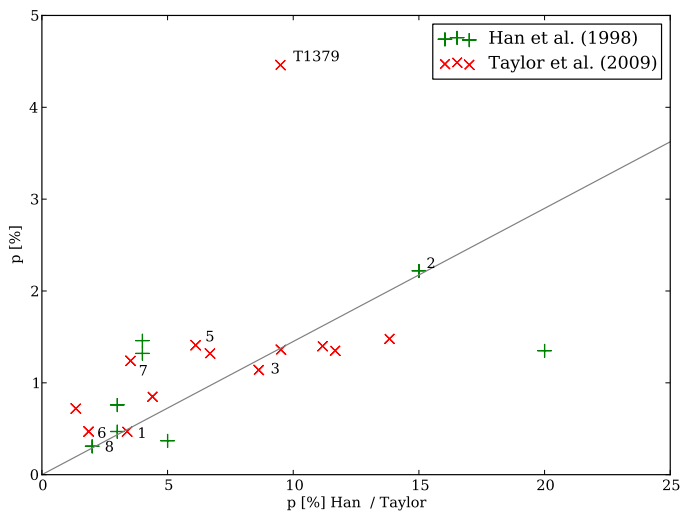}
\caption[Comparison of polarization degrees at 90~cm with literature values.]{Measured polarization degree from the Faraday cube at 90~cm versus the values measured by \cite{han} and \cite{taylor} at 20~cm. The solid grey line is a linear fit through all points ($DP=0.14\pm0.02$, $\chi^2=0.76$).}
\label{pvsp}
\end{center}
\end{figure}

It might be possible that the deviating sources in Fig.~\ref{RMcompare} are of a different type from those near the solid line. If these sources suffer from internal Faraday depolarization, the intrinsic FD at low frequencies can be entirely different from that at high frequencies, since the visible region of polarized emission becomes smaller (because all the emission from deeper layers is depolarized and will no longer contribute to the emission and Faraday rotation). In that case these sources should be more strongly depolarized than others. In Figure~\ref{pvsp} the polarization degree measured at 90~cm is plotted against that measured at 20~cm by \cite{han} and \cite{taylor}. The deviating sources are denoted with the same numbers as given in Fig~\ref{RMcompare}. The solid grey line is a linear fit through all points; its slope corresponds the mean depolarization between 90~cm and 20~cm ($DP(90,20)=0.14\pm0.02$, $\chi^2=0.76$). Taking into account the large scatter, all sources (except one) follow the mean depolarization. In particular, the sources deviating in Fig.~\ref{RMcompare} are not systematically more strongly depolarized.

The strong depolarization of the background sources could originate (1) in the Galactic foreground, (2) in M31, (3) in intervening galaxies along the line of sight, or (4) in the sources themselves. The condition for model (1) is that the angular extent of the source is larger than that of a typical turbulence cell. None of our sources shows significant extension in total intensity on the scale of our telescope beam (4\arcmin), which is much smaller than the angular size of a turbulence cell in the Galactic foreground of typically 50~pc linear size out to several kiloparsecs within the Galaxy. Model (2) can be excluded because the amount of depolarization does not increase towards the inner part of M31. We conclude that the depolarization occurs in distant intervening galaxies or within the sources.

Similarly, the multiple components seen in the Faraday spectra of Figs.~\ref{rm_spec1}--\ref{rm_spec6} can hardly originate in the Galactic foreground, because they would have to emit in polarization on similar levels as the background sources themselves (Fig.~\ref{fig:rmsynth}). They are either intrinsic features of these sources, for instance, the lobes of radio galaxies, or occur in the turbulent medium of intervening galaxies \citep{bernet12}. If the source is covered by a discrete number of turbulent cells in the intervenor or in the lobe, a corresponding number of components appears in the Faraday spectrum (see Fig.~5 in \cite{bernet12}). If unresolved, these components lead to depolarization. If the number of components is large, depolarization can be described by dispersion in Faraday rotation (see equation (\ref{DP_S}) below).

This scenario is supported by the fact that the two sources with least depolarization (T1379 and 37W014/T1119) have simple Faraday spectra without multiple components outside the range of $\pm30$~rad~m$^{-2}$, which is affected by polarized emission from the nearby foreground and by instrumental polarization. Furthermore, the three sources with the highest degrees of polarization (T1379, 37W154, and 37W019) also have simple Faraday spectra. A more detailed analysis is hampered by the small number of sources and by the fact that for most sources no data (distance, optical classification, and spectrum) are available from the literature.

37W115/T1237 is a known AGN, consisting of three components (core and two lobes, \citealt{morgan13}) that are unresolved at the resolution of our 90~cm data. It is possible that at low frequencies one of the lobes becomes depolarized, and in turn the polarized emission of the other lobe dominates (the Laing--Garrington effect, \citealt{laing-garrington}). In this case a comparison with 20~cm data is not practical, since one essentially observes two different source patterns. The same may also be the case for radio galaxies at larger distances.

Finally, none of our sources fits into the depolarization model for compact, steep-spectrum (CSS) sources by \cite{rossetti08}. This model predicts that the degree of polarization should be constant at long wavelengths. The strong depolarization seen in Fig.~\ref{pvsp} indicates that our sources are not of type CSS.

\subsection{Diffuse polarized emission from M31 and depolarization at 90~cm}\label{M31pol}

No extended polarized emission of M31 is visible by eye in the Faraday cube. Internal Faraday dispersion is the dominating depolarization mechanism towards low frequencies. Changes of the polarization angle increase with decreasing frequency. Due to turbulent cells in the magnetized plasma, the polarized emission becomes increasingly patchy and the polarization angle randomized, which is the true cause of the depolarization.

Thus it is unclear whether any coherent polarized emission on large scales can be expected at all (either in the image-plane, or in FD), therefore we propose a new method for uncovering weak diffuse polarized emission at low frequencies.

From observations at GHz frequencies it is known that the polarized emission in M31 is strongest around the 10-kpc ring, showing an RM range of roughly -200~rad/m$^2$ to 0~rad/m$^2$. The intrinsic RM in M31 is thus -100~rad/m$^2$ to +100~rad/m$^2$, shifted by the RM of the Galactic foreground by $\sim-93$~rad/m$^2$. RM synthesis allows us to compare the positive and negative FD ranges. An excess signal in the negative FD range, which is confined to the position of M31, would be a detection of M31 in polarization.

We used the range of $(-100\pm50)$~rad/m$^2$ and compared it with the complementary range of $(+100\pm50)$~rad/m$^2$. The selected range is thus centred on the FD where we expect most of the signals. Since we are in the Faraday-thick regime (\emph{opaque layer approximation}, see e.g. \citealt{sokoloff98}), the total range in rotation measure will be smaller. The restriction to $(-100\pm50)$~rad/m$^2$ also excludes the range in the Faraday cube that is affected by foreground emission and instrumental polarization (see end of Section~\ref{catalogue}).

For each pixel in the map, we integrated along the Faraday spectrum over the absolute values in the selected positive FD range and subtracted that result from the integral along the Faraday spectrum in the selected negative FD range. Since we integrated over the amplitude of a complex-valued function, the result is by definition positive and contains a positive non-zero component from the noise. The second component in the spectrum consists of signals from M31 itself plus a few weak, polarized point sources.

Owing to the rotation measure of the Galactic foreground, the positive FD range does not contain any significant signals of the second component. By subtracting the complementary positive FD range, we subtracted the contribution of the noise, which is equal over the entire spectrum. The residual is therefore an estimate for the intrinsic polarized signal from M31. The positive bias from the noise component is removed with the subtraction, but due to the nonlinear addition of noise and signals in polarized intensity, weak signals are suppressed by our subtraction method, and our result is an underestimate.

\begin{figure}[h]
\begin{center}
\includegraphics[trim=30px 0 0 0px,clip,scale=0.55]{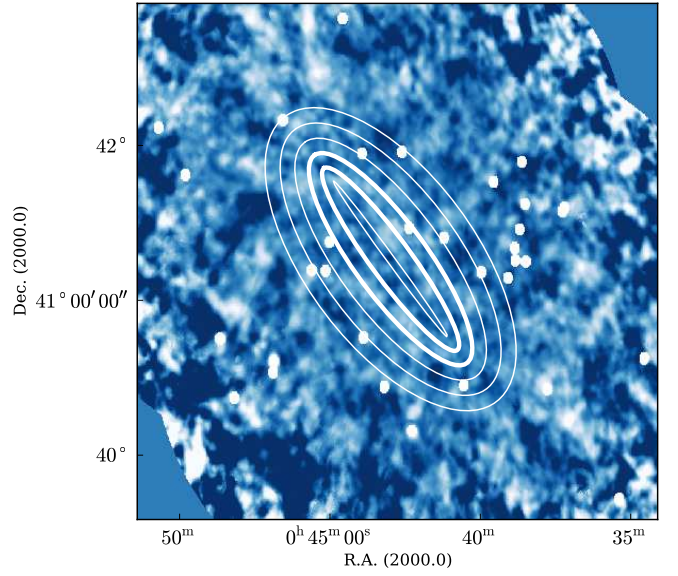}
\caption{Residual polarization map after integrating each pixel over the FD ranges $(-100\pm50)$~rad/m$^2$ and $(+100\pm50)$~rad/m$^2$ and subtracting the results. Overlaid are the outlines of the ellipses used to integrate in the image plane. The outline of the ellipse on the 10-kpc ring is marked in boldface. Detected point sources have been masked (seen as white points).}
\label{peakmap}
\end{center}
\end{figure}

Figure~\ref{peakmap} shows the resulting residual map. Since there is still no clear detection of polarized emission, we integrated over the residual map in the image plane. After masking the detected point sources, we defined five ellipses with a width of~$6'$: one on the exact position where the 10-kpc ring is seen in total power (and where we expect the strongest polarization signal), one inside the ring, and three outside the ring. The outlines of the ellipses are overlaid in Figure~\ref{peakmap} and the results are listed in Table~\ref{tab:integflux}. The error in each ellipse was estimated by the standard deviation of the value in each ellipse multiplied by the sqare-root of independent resolution elements.

\begin{table}
\begin{center}
\begin{tabular}{rcccc}
Ellipse & \#$_{\text{resol.elem.}}$ & I [mJy] & PI [mJy] & p [\%]\\ \hline
inner   & $\sim35$ & $2567 \pm 36$  & $3\pm3$ & $0.12\pm0.12$ \\
ring    & $\sim45$ & $4100 \pm 19$  & $9\pm3$ & $0.22\pm0.07$\\
outer   & $\sim54$ & $2325 \pm 22$  & $7\pm3$ & $0.30\pm0.13$\\ \hline
sum     & & $8992 \pm 77$  & $19\pm5$ & $0.21\pm0.05$\\ \hline\hline
outer2  & $\sim63$ &  $876 \pm 21$  & $-4\pm3$ & \\
outer3  & $\sim72$ &  $201 \pm 18$  & $-4\pm4$ & \\ \hline
\end{tabular}
\caption{Number of independent resolution elements, integrated total flux density, residual integrated polarized flux density over the selected FD range (see text) and degree of polarization. Line 5 gives the total of inner-, ring- and outer ellipse and the resulting polarization degree. Integration is made over residual histograms (see text), which explains the occurrence of negative values.}
\label{tab:integflux}
\end{center}
\end{table}

\begin{figure}[h]
\begin{center}
\includegraphics[scale=0.4]{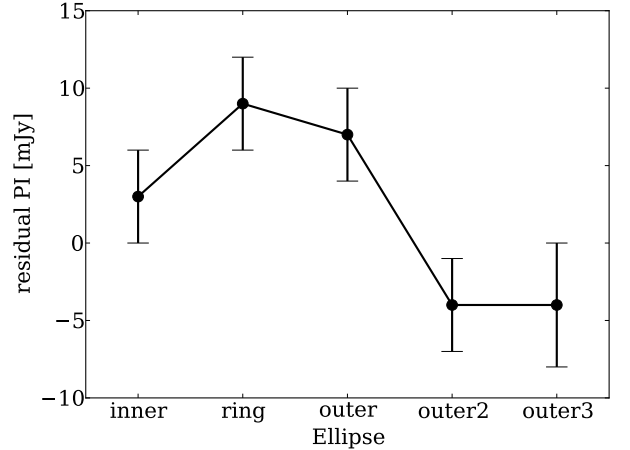}
\caption{Integrated residual polarized flux densities for the FD range $\pm100\pm50$~rad~m$^{-2}$ for the different ellipses (see Table~\ref{tab:integflux}).}
\label{fig:FDranges}
\end{center}
\end{figure}

We repeated the analysis for the complementary FD ranges $\pm200\pm50$~rad~m$^{-2}$ (not shown here). As expected, no excess of polarized emission was found for these FD ranges.

Since the excess is confined to the position of polarized emission at GHz frequencies and to the expected FD range, we identified it as polarized emission originating from M31.
However, the weakness of the signals does not allow us to measure the distribution of the FD along the ring and compare this to previous results. It must be somewhat uniform or it would be visible at particular azimuthal ranges in the residual map in Figure~\ref{peakmap}. The layer of the regular field known to exist from observations at higher
frequencies is broken up by Faraday dispersion into small regions emitting in polarization. This is clearly visible in the PI map at 20~cm by \cite{vla20}. Instead of one broad structure in the Faraday spectrum, we expect many components. Our integration collects all these features.

Now the depolarization between 6~cm and 90~cm can be estimated. The 6~cm polarized intensity map (\citealt{phd}, Gie\ss\"ubel et al. in prep.) was smoothed to the resolution of the 90~cm map and the total polarized flux density within the inner three ellipses was calculated ($PI_{6}\approx0.296$~Jy). The depolarization can then be calculated as in \cite{beck07},
\begin{equation}
DP(90,6)=(PI_{90}/PI_6)*(\nu_6/\nu_{90})^{\alpha_n},
\end{equation}
where $\alpha_n=-1.0$ is the synchrotron spectral index and $\nu_i$ the respective frequency, leading to $DP(90,6)=0.005\pm0.002$.

The depolarization caused by internal Faraday dispersion between 20~cm and 6~cm was determined by \cite{fletcher04} to be $DP(20,6)\approx0.1$. Referring to \cite{burn}, we can calculate the expected depolarization at 90~cm.
At these low frequencies depolarization is dominated by internal Faraday dispersion.
\begin{equation}
DP_{int}=\frac{1-\exp(-S)}{S}\label{DP_int}
\end{equation}
with
\begin{equation}
S=2\sigma_{RM}^2\lambda^4 \label{DP_S}.
\end{equation}
The factor $\sigma_{RM}$ describes the dispersion of the medium in~rad~m$^{-2}$. In a simplified model of a magneto-ionic medium it can be written as \citep{arshakian11}
\begin{align}
\sigma_{RM}^2&=(0.81\, n_e\, B_{turb}\, d)^2\,\frac{fL}{d}\notag \\
&\simeq(0.81 \langle n_e\rangle \langle B_{turb}\rangle)^2\,\frac{Ld}{f}\label{DP_sigma},
\end{align}
where $n_e=\langle n_e\rangle/f$ is the thermal electron density in cm$^{-3}$ within the turbulent cells of size $d$ in pc, $\langle n_e\rangle$ is the average electron density in the volume along the pathlength traced by the telescope beam in pc, $f$ is the filling factor of the cells, $\langle B_{turb}\rangle$ the strength of the turbulent field in $\mu$G and $L$ the pathlength along the line of sight through the Faraday-active emitting layer. The term in brackets describes the rotation measure of a single turbulent cell, while $\frac{fL}{d}$ gives the number of cells along the line of sight.

The following quantities were used: $\langle n_e\rangle=0.015$~cm$^{-3}$ (in concordance with the depolarization caused by internal Faraday dispersion of 0.1, 0.3 and 0.2 measured by \cite{fletcher04} for different rings at $8-10$~kpc, $10-12$~kpc and $12-14$~kpc, respectively), a filling factor of $f=0.2$ \citep{walterbos94}, $B_{turb}=5~\mu$G and $d=50$~pc \citep{fletcher04}. According to \cite{fletcher04}, the scale height of the synchrotron-emitting layer is $h_{syn}=300$~pc. The scale height was measured from the mid-plane. The pathlength through the entire layer along the line of sight is thus
\begin{equation}
L=\frac{2 h_{syn}}{\cos(i)}\approx2300 \text{ pc},
\end{equation}
where $i=75\grad$ \citep{chemin} is the inclination angle of the disk.

The classical formula by \cite{burn} would yield $DP(90,6)\approx0.0004$, $p\approx0.03$\%, which essentially means total depolarization. In the model by \cite{tribble} the wavelength dependence turns from $\lambda^4$ to $\lambda^2$ for longer wavelengths and equation~(\ref{DP_S}) becomes
\begin{equation}
S=2\sigma_{RM}\lambda^2
\end{equation}
(see also \citealt{sokoloff98,arshakian11}). This is because the dispersion causes the (spatial) correlation length of the polarized emission to decrease with increasing wavelength, until it drops below the size of the turbulent cells. In this picture, no extended structures would be visible from M31 in the Faraday cube, which is consistent with our observations.

Since $S\gg1$, eq.~\ref{DP_int} becomes
\begin{equation}
DP=\frac{1}{2\sigma_{RM}\lambda^2}\label{DP_l2},
\end{equation}
which (using the values given above) results in $DP=0.015$, $p=1.09$\%.

The measured values of $DP(90,6)=0.005\pm0.002$, $p=0.21\pm0.05$\% are thus between the predictions of Burn and Tribble (but closer to the latter, which is just outside our $3\sigma$ uncertainty bound). Our detection is a lower limit for the true polarized signal. If some polarized emission from M31 is extended in Faraday space, we underestimate the detected polarized flux even more because of missing zero frequencies \citep[see][]{brentjens05}. We are therefore unable to comment on the validity of the Tribble formula, but we can confidently rule out a $\lambda^4$ dependence of the depolarization for low frequencies.

\section{Conclusions}

We have presented the first detection of polarized emission from a nearby galaxy at 90~cm. The polarized emission is mainly confined to the 10-kpc ring. At these low frequencies the emission comes from cosmic-ray electrons with lower energies. They suffer less from energy losses and therefore are able to propagate farther out from the disk or into a halo than at higher frequencies. However, no signs of a radio halo can be seen in our total power map (Fig.~\ref{M31_wsrt}). The reason may be the preferred propagation of cosmic-ray electrons along the highly ordered field in the ring and suppression of diffusion perpendicular to the ring \citep{fletcher04}.

Depolarization by internal Faraday dispersion becomes strong at long wavelengths. Our observations show that the strong $\lambda^4$ wavelength dependence predicted by \cite{burn} underestimates the remaining polarization at 325 MHz. We were unable to conclude whether a $\lambda^2$ depolarization law as predicted by Tribble (1991) is correct, but we note that the corresponding DP does fall within our $3\sigma$ error margin. The amount of depolarization furthermore depends on the magnetic field strength, thermal electron density, and the pathlength along the line of sight and can accordingly be very different in other galaxies, depending on the conditions. A detection of diffuse polarized emission from nearby galaxies with LOFAR will be difficult and requires very high sensitivity, achievable with the RM synthesis technique in combination with a broad bandwidth. With the same parameters in eq.~(\ref{DP_l2}) as used in Sect.~\ref{M31pol}, depolarization in the LOFAR high band around 150~MHz will be about ten times stronger. Searches for diffuse polarized emission in galaxies with LOFAR are most likely to yield detections in outer disks and halos, where $\sigma_{RM}$ is expected to be much smaller than in the ring of M31.

Using polarized background sources to probe the magnetic fields of nearby galaxies as a foreground screen seems a more promising prospect. Hence, we compiled a catalogue of all polarized point-like sources. The RM synthesis analysis resulted in 33 detections, but fewer than 50\% of the sources listed by \cite{han} at 20~cm were detected at 90~cm. Fig.~\ref{rm_map} shows that the number of detected sources is not sufficient to provide a grid of sources for probing the magnetic field of M31 as a foreground screen to these sources. According to \cite{stepanov}, for a galaxy such as M31, about 20 polarized sources on a cut along the projected minor axis would be needed to detect the dominant field structure (the number depends on the inclination of the galaxy, for an inclination of $45\grad$ a total number of ten would be sufficient), but the 33 sources detected here are scattered over the entire field of view. Furthermore, the estimate by \cite{stepanov} assumed a constant contribution of the Milky Way foreground, which is not the case here (see Section~\ref{catalogue}). In conclusion, the number of polarized sources from our 90~cm observations is insufficient.

A higher angular resolution will help with the detection of more sources, since
fewer features extended in RA/DEC (blending with the emission
of sources) will be present in the Faraday cube. LOFAR can provide the
necessary resolution, both angular and in FD.

The analysis in Section \ref{compar} showed that a more detailed knowledge about the structure of the sources is necessary to interpret their Faraday spectra. About 30\% of the sources from the literature showed a systematically different FD at 90~cm than at 20~cm. Additional studies are needed to understand the cause of this deviation. A possible explanation is the presence of radio lobes, as discussed in Sect.~\ref{compar}. The systematics of the deviation (see Fig.~\ref{RMcompare} in Sect.~\ref{compar}) indicates that it may be possible to define distinct classes of sources and select those suitable for a background grid for future observations with LOFAR. The depolarization of the sources is on average $DP(90,20)=0.14\pm0.02$, more than seven times stronger at 90~cm than at 20~cm. If depolarization occurs due to a limited number of turbulent cells in the radio lobes or in an intervening galaxy, these cells appear as multiple components at fixed FDs in the Faraday spectrum and should be clearly resolvable with LOFAR \citep{beck12}. On the other hand, an emitting and Faraday-rotating region with a regular magnetic field on the line of sight, such as a radio lobe, an intervening galaxy, or a nearby region in the Milky Way, generates an extended feature in the Faraday spectrum that can be recognized at high frequencies or as a double peak at low frequencies. Future studies should combine data from GHz and MHz frequencies with RM synthesis.

\begin{acknowledgements}
The authors would like to thank Björn Adebahr and Carlos Sotomayor (Ruhr-Universität Bochum) for their initial help with calibration and peeling in \texttt{CASA}, and Elly M. Berkhuijsen for valuable comments.

We are also grateful for the comments of the anonymous referee, which helped to improve our paper.

RB acknowledges support from DFG FOR1254.

TGA acknowledges support by DFG project number Os  177/2-1.

The presented results and images have in part been published in \cite{phd}.

The Westerbork Synthesis Radio Telescope is operated by the ASTRON (Netherlands Institute for Radio Astronomy) with support from the Netherlands Foundation for Scientific Research (NWO).
\end{acknowledgements}

\bibliographystyle{aa} 
\bibliography{mybib.bib} 

\clearpage
\begin{appendix}
\onecolumn
\section{Catalogue of detected sources}
\begin{table}[ht]
\begin{sideways}
\begin{minipage}[t]{23.1cm}
\small
\begin{tabular}{|c|cc|ccc|ccc|ccc|}\hline
Name & \textsc{Ra} & \textsc{Dec} & $\phi$ [rad/m$^2$]& PI [mJy] & p [\%] & RM [rad/m$^2$] & PI [mJy] & p [\%] &  RM [rad/m$^2$] & PI [mJy] & p [\%] \\
& & & & & & \multicolumn{3}{c|}{\citep{han}} & \multicolumn{3}{c|}{\citep{taylor}} \\ \hline
J005058+420637 & 00h50m58.2s &   42d 6m37.1s & 28.0 $\pm$ 0.3 & 4.18 $\pm$ 0.15 & 1.63 &     &     &   &     &     &    \\
J005000+414836 & 00h50m00.8s &   41d48m36.0s & -94.0 $\pm$ 0.6 & 2.36 $\pm$ 0.15 & 1.73 &     &     &   &     &     &    \\
T1421 & 00h48m43.9s &   40d45m36.7s & -86.0 $\pm$ 0.2 & 9.12 $\pm$ 0.15 & 1.48 &     &     &   & -81.9 $\pm$ 1.6 & 23.46 $\pm$ 0.18 & 13.83 $\pm$ 0.11\\
T1407$^{(5)}$ & 00h48m13.0s &   40d22m57.2s & -42.0 $\pm$ 0.1 & 26.38 $\pm$ 0.15 & 1.41 &     &     &   & -96.7 $\pm$ 1.8 & 28.28 $\pm$ 0.23 & 6.12 $\pm$ 0.05\\
T1375? & 00h46m56.1s &   40d32m54.5s & -94.0 $\pm$ 0.5 & 3.03 $\pm$ 0.15 & 1.36 &     &     &   & -88.2 $\pm$ 12.9 & 4.95 $\pm$ 0.37 & 9.51 $\pm$ 0.71\\
T1379? & 00h46m55.1s &   40d37m24.6s & -86.0 $\pm$ 0.1 & 10.02 $\pm$ 0.15 & 4.46 &     &     &   & -91.3 $\pm$ 14.4 & 5.0 $\pm$ 0.39 & 9.49 $\pm$ 0.74\\
37W235?T1373$^{(1)}$ & 00h46m42.2s &   42d10m11.9s & -104.0 $\pm$ 0.9 & 1.61 $\pm$ 0.15 & $\downarrow$0.47 &     &     &   & -199.9 $\pm$ 119 & 4.56 $\pm$ 0.26 & 3.39 $\pm$ 0.19\\
37W219 & 00h45m40.2s &   41d12m334s & -72.0 $\pm$ 0.8 & 1.87 $\pm$ 0.15 & 1.46 & -83 $\pm$ 9 & 0.5 $\pm$ 0.05 & 4 &     &     &    \\
37W211/T1332 & 00h45m12.3s &   41d12m20.8s & -86.0 $\pm$ 0.8 & 1.81 $\pm$ 0.15 & 1.35 & -86 $\pm$ 1 & 4.2 $\pm$ 0.06 & 20 & -85.9 $\pm$ 15.4 & 3.61 $\pm$ 0.29 & 11.67 $\pm$ 0.94\\
37W207B$^{(2)}$ & 00h45m04.7s &   41d23m21.4s & -50.0 $\pm$ 0.4 & 3.99 $\pm$ 0.15 & 2.22 & -129 $\pm$ 7 & 0.55 $\pm$ 0.1 & 15 &     &     &    \\
J004440+424938 & 00h44m40.6s &   42d49m38.9s & -68.0 $\pm$ 0.4 & 3.82 $\pm$ 0.15 & $\downarrow$0.27 &     &     &   &     &     &    \\
37W175b & 00h43m58.7s &   41d57m40.3s & -124.0 $\pm$ 0.9 & 1.67 $\pm$ 0.15 & $\downarrow$0.37 & -127 $\pm$ 3 & 2.09 $\pm$ 0.06 & 5 &     &     &    \\
37W172$^{(8)}$ & 00h43m54.7s &   40d46m40.3s & -150.0 $\pm$ 1.1 & 1.25 $\pm$ 0.15 & $\downarrow$0.31 & -68 $\pm$ 3 & 2.7 $\pm$ 0.14 & 2 &     &     &    \\
37W154? & 00h43m12.3s &   40d27m41.3s & -22.0 $\pm$ 0.2 & 6.12 $\pm$ 0.15 & 21.62 &     &     &   &     &     &    \\
37W131 & 00h42m36.6s &   41d58m11.7s & -90.0 $\pm$ 1.0 & 1.43 $\pm$ 0.15 & $\downarrow$0.76 & -93 $\pm$ 2 & 1.88 $\pm$ 006 & 3 &     &     &    \\
37W115/T1237?$^{(6)}$ & 00h42m22.0s &   41d28m56.5s & -30.0 $\pm$ 0.3 & 4.95 $\pm$ 0.15 & $\downarrow$0.47 & -92 $\pm$ 1 & 8.58 $\pm$ 0.06 & 3 & -71.1 $\pm$ 10.4 & 6.82 $\pm$ 0.34 & 1.86 $\pm$ 0.09\\
37W114 & 00h42m17.3s &   40d10m41.2s & -52.0 $\pm$ 0.2 & 8.22 $\pm$ 0.15 & $\downarrow$0.39 &     &     &   &     &     &    \\
37W074A/B$^{(4)}$ & 00h41m11.4s &   41d25m09.2s & -54.0 $\pm$ 0.7 & 2.04 $\pm$ 0.15 & 2.05 & -105 $\pm$ 5 &     &   &     &     &    \\
37W058 & 00h40m33.3s &   40d28m06.7s & 28.0 $\pm$ 0.6 & 2.22 $\pm$ 0.15 & 3.65 &     &     &   &     &     &    \\
37W045/T1126 & 00h39m55.9s &   41d11m48.8s & -100.0 $\pm$ 0.5 & 2.72 $\pm$ 0.15 & 1.32 & -113 $\pm$ 6 & 0.55 $\pm$ 0.06 & 4 & -103.2 $\pm$ 13.9 & 4.63 $\pm$ 0.3 & 6.69 $\pm$ 0.43\\
37W033 & 00h39m28.9s &   41d46m31.2s & -122.0 $\pm$ 0.6 & 2.38 $\pm$ 0.15 & $\downarrow$0.86 &     &     &   &     &     &    \\
J003901+410928 & 00h39m01.6s &   41d 9m28.0s & -32.0 $\pm$ 0.5 & 2.91 $\pm$ 0.15 &  &     &     &   &     &     &    \\
J003847+412056 & 00h38m47.6s &   41d20m56.3s & -36.0 $\pm$ 0.5 & 2.76 $\pm$ 0.15 &  &     &     &   &     &     &    \\
37W021 & 00h38m46.6s &   41d16m11.1s & -62.0 $\pm$ 1.0 & 1.48 $\pm$ 0.15 & $\downarrow$0.99 &     &     &   &     &     &    \\
J003825+411538 & 00h38m25.3s &   41d15m38.2s & -36.0 $\pm$ 0.6 & 2.3 $\pm$ 0.15 &  &     &     &   &     &     &    \\
J003830+415409 & 00h38m30.8s &   41d54m09.2s & -42.0 $\pm$ 0.3 & 5.35 $\pm$ 0.15 & 3.39 &     &     &   &     &     &    \\
37W019? & 00h38m36.5s &   41d28m09.8s & -38.0 $\pm$ 0.4 & 3.19 $\pm$ 0.15 & 8.8 &     &     &   &     &     &    \\
37W014/T1119 & 00h38m25.1s &   41d37m53.3s & -110.0 $\pm$ 0.3 & 4.26 $\pm$ 015 & $\downarrow$0.72 &     &     &   & -94.1 $\pm$ 6.9 & 8.78 $\pm$ 0.28 & 1.35 $\pm$ 0.04\\
37W008 & 00h37m45.1s &   40d26m16.7s & -32.0 $\pm$ 0.3 & 5.26 $\pm$ 0.15 & 1.79 &     &     &   &     &     &    \\
T1067 & 00h37m09.1s &   41d34m56.1s & -96.0 $\pm$ 0.4 & 3.52 $\pm$ 0.15 & 1.4 &     &     &   & -91.7 $\pm$ 14.4 & 4.29 $\pm$ 0.28 & 11.17 $\pm$ 0.73\\
T1061$^{(3)}$ & 00h37m05.0s &   41d36m10.3s & -48.0 $\pm$ 0.4 & 3.85 $\pm$ 0.15 & 1.14 &     &     &   & -107.6 $\pm$ 12.8 & 4.81 $\pm$ 0.29 & 8.63 $\pm$ 0.52\\
T1006 & 00h35m25.2s &   39d43m32.2s & -102.0 $\pm$ 0.4 & 3.37 $\pm$ 0.15 & $\downarrow$0.85 &     &     &   & -110.8 $\pm$ 11.4 & 5.86 $\pm$ 0.29 & 4.4 $\pm$ 0.22\\
T974$^{(7)}$ & 00h34m29.2s &   40d37m34.2s & -104.0 $\pm$ 0.3 & 5.17 $\pm$ 0.15 & 1.24 &     &     &   & -58.2 $\pm$ 14.5 & 4.15 $\pm$ 0.31 & 3.53 $\pm$ 0.26\\ \hline
\end{tabular}
\caption{Catalogue of the detected Faraday sources. Numbers in brackets behind the names ($^{(1)(2)(3)(4)(5)(6)(7)(8)}$) are sources with deviating values with regard to the literature, see Section~\ref{compar} and Figures~\ref{RMcompare} and~\ref{pvsp}. J003901+410928, J003847+412056 and J003825+411538 from the triplet (see Section~\ref{catalogue}) a peculiar foreground pattern with no counterpart in Stokes I (hence there are no values for $p$ given). Values of $p$ marked with $\downarrow$ denote upper limits, assuming a reasonable calibration limit of $p=1$\%. The sources are ordered from east to west.}\label{RMcat}
\end{minipage}
\end{sideways}
\end{table}

\section{Faraday spectra of the sources detected at $\lambda$90~cm}\label{spectra.app}

The following pages show the Faraday spectra of the sources detected at 90~cm. We refer to Section~\ref{catalogue} for details.
The peak associated with the detected source is marked with a blue arrow and a blue error bar. Green error bars and dashed lines show literature values from \cite{han}, the red errorbars and dashed lines from \cite{taylor} where available.

An important selection criterion for a source to be included in the catalogue was a clear recognition as a point-like source in the cube (in \textsc{Ra}/\textsc{Dec}). This is why sometimes higher peaks than the one marked as detection are present in the spectra. Any other peaks that possibly appear in the spectra are partly extended features.

The grey area marks the range of FD ($\pm30$~rad~m$^{-2}$) that probably is affected by emission from the nearby foreground of the Milky Way and by instrumental polarization.

Like in Table~\ref{RMcat}, the sources are ordered from east to west.

\begin{figure}[hp!]
\begin{center}
\includegraphics[scale=0.45,angle=90]{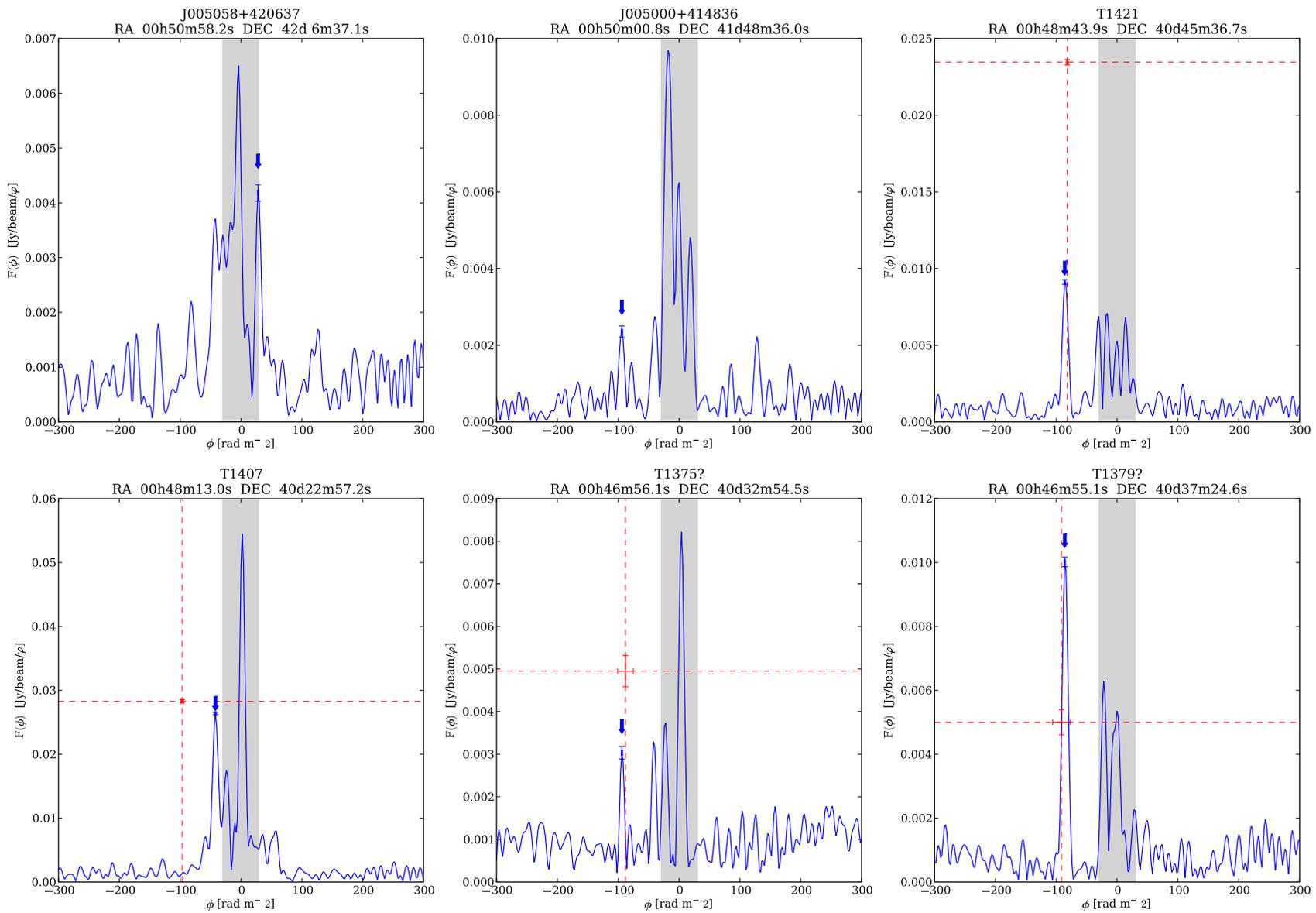}
\caption{Faraday Spectra of the detected sources. See text on previous page for details.}
\label{rm_spec1}
\end{center}
\end{figure}
\begin{figure}[hp!]
\begin{center}
\includegraphics[scale=0.45,angle=90]{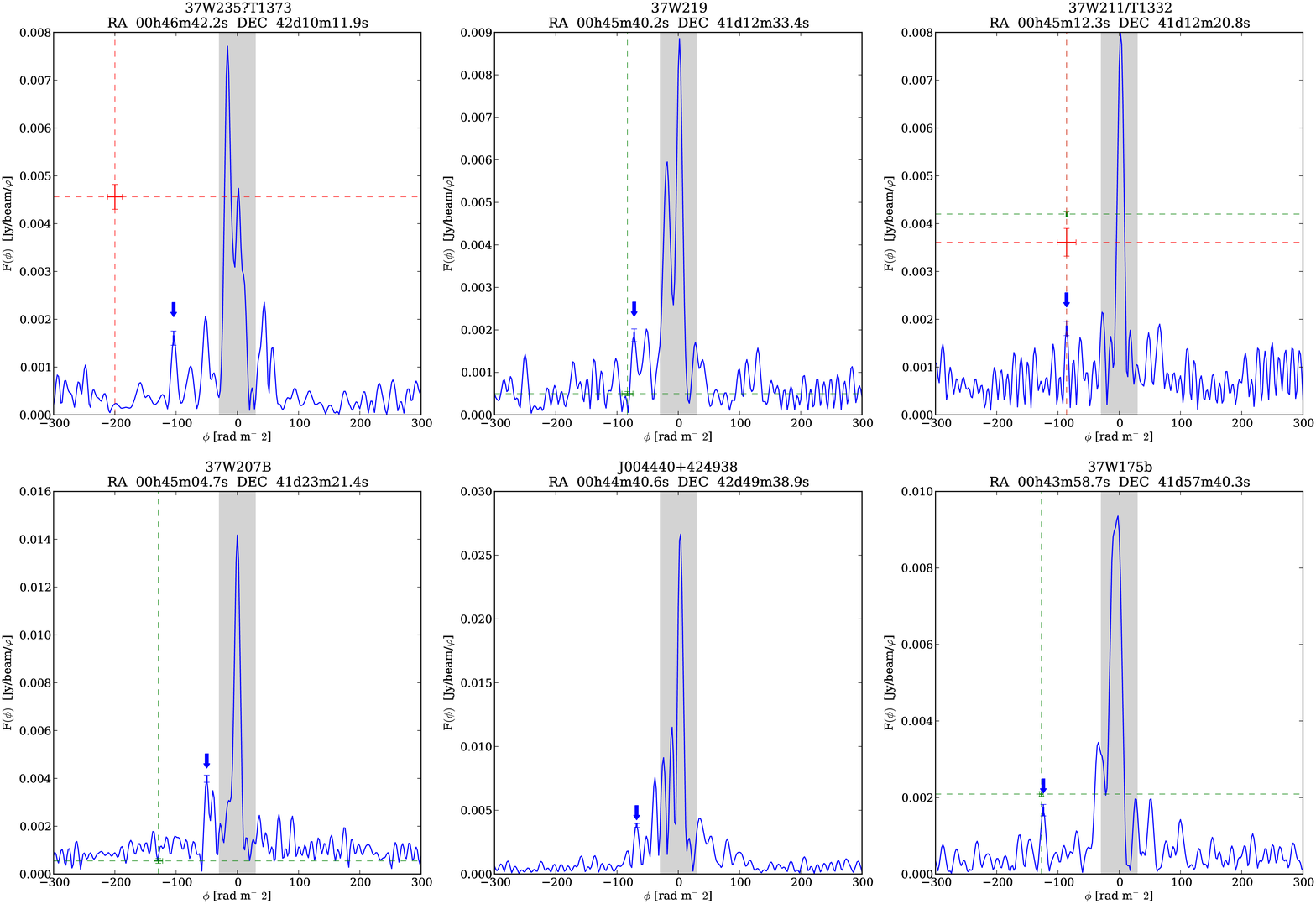}
\caption{Faraday Spectra of the detected sources (continued).}
\label{rm_spec2}
\end{center}
\end{figure}
\begin{figure}[hp!]
\begin{center}
\includegraphics[scale=0.45,angle=90]{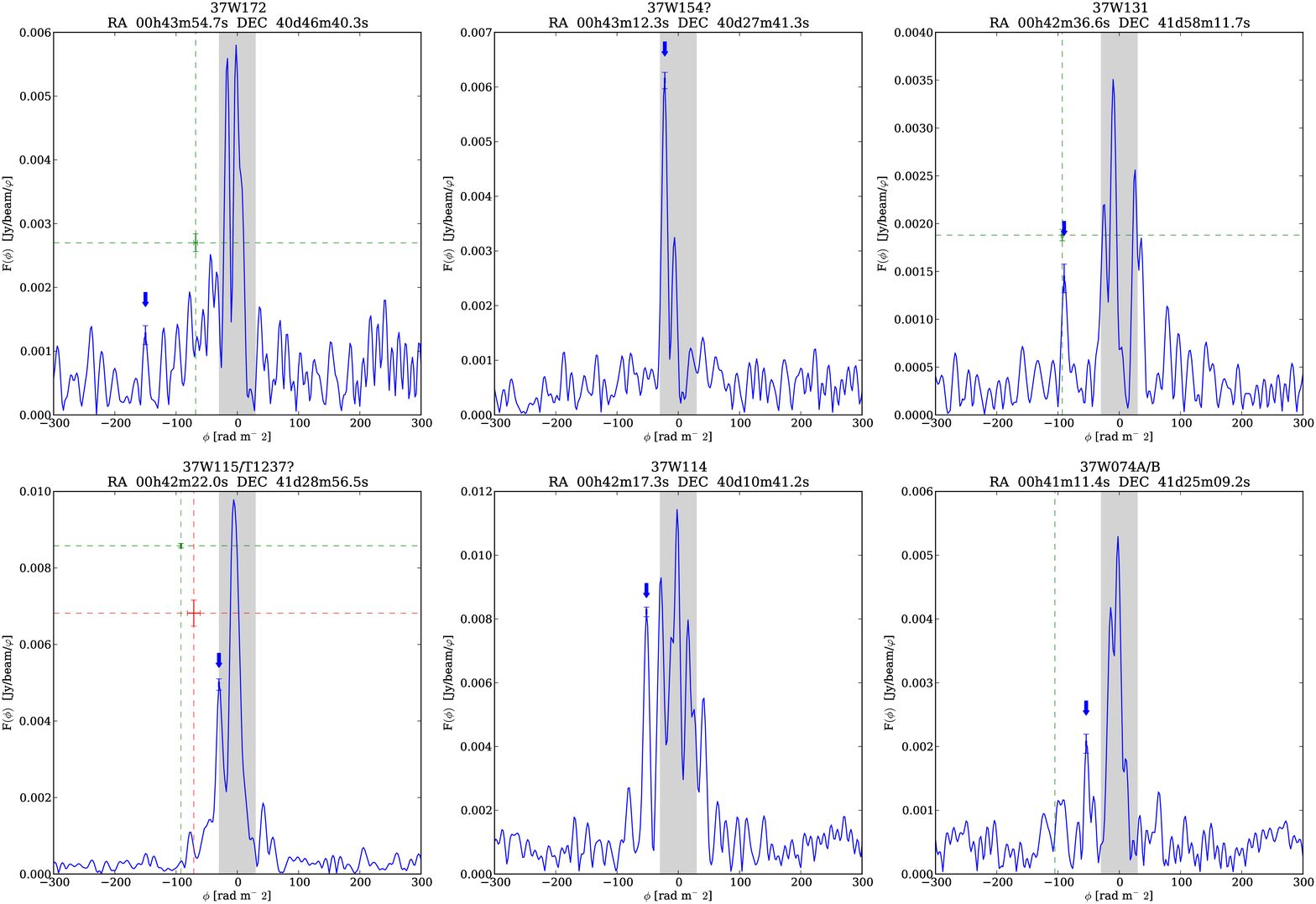}
\caption{Faraday Spectra of the detected sources (continued).}
\label{rm_spec3}
\end{center}
\end{figure}
\begin{figure}[hp!]
\begin{center}
\includegraphics[scale=0.45,angle=90]{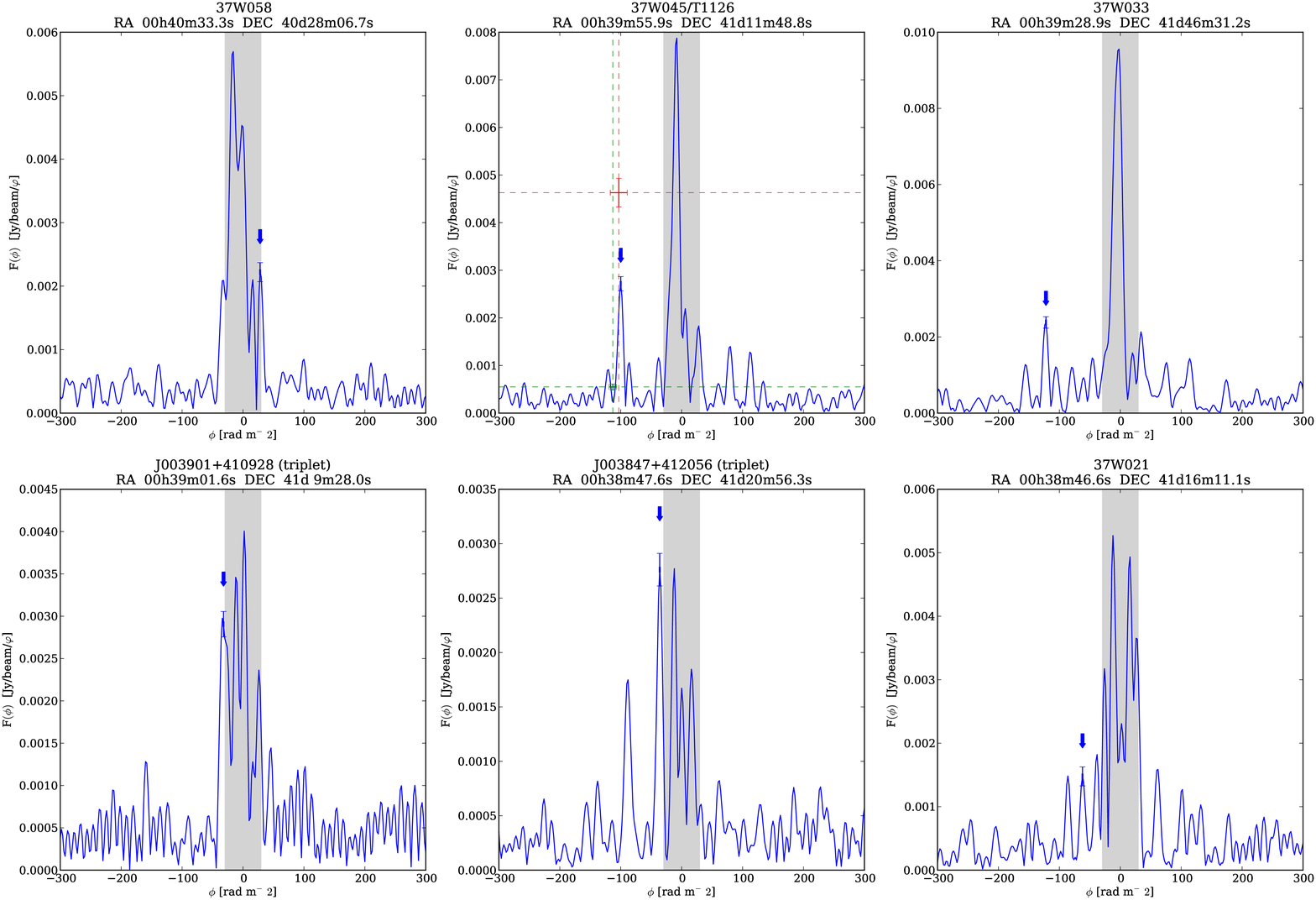}
\caption{Faraday Spectra of the detected sources (continued).}
\label{rm_spec4}
\end{center}
\end{figure}
\begin{figure}[hp!]
\begin{center}
\includegraphics[scale=0.45,angle=90]{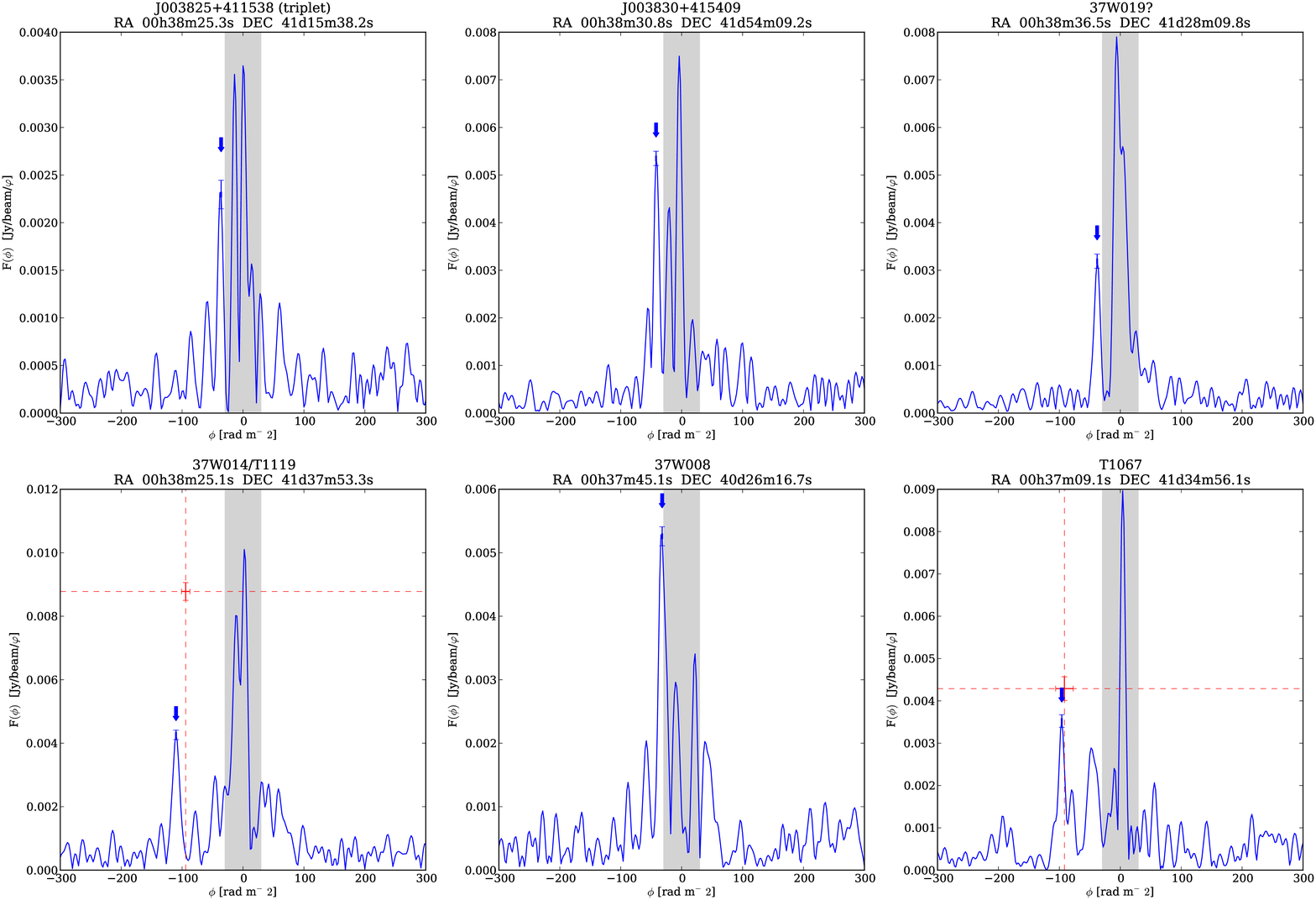}
\caption{Faraday Spectra of the detected sources (continued).}
\label{rm_spec5}
\end{center}
\end{figure}
\begin{figure}[hp!]
\begin{center}
\includegraphics[scale=0.45,angle=90]{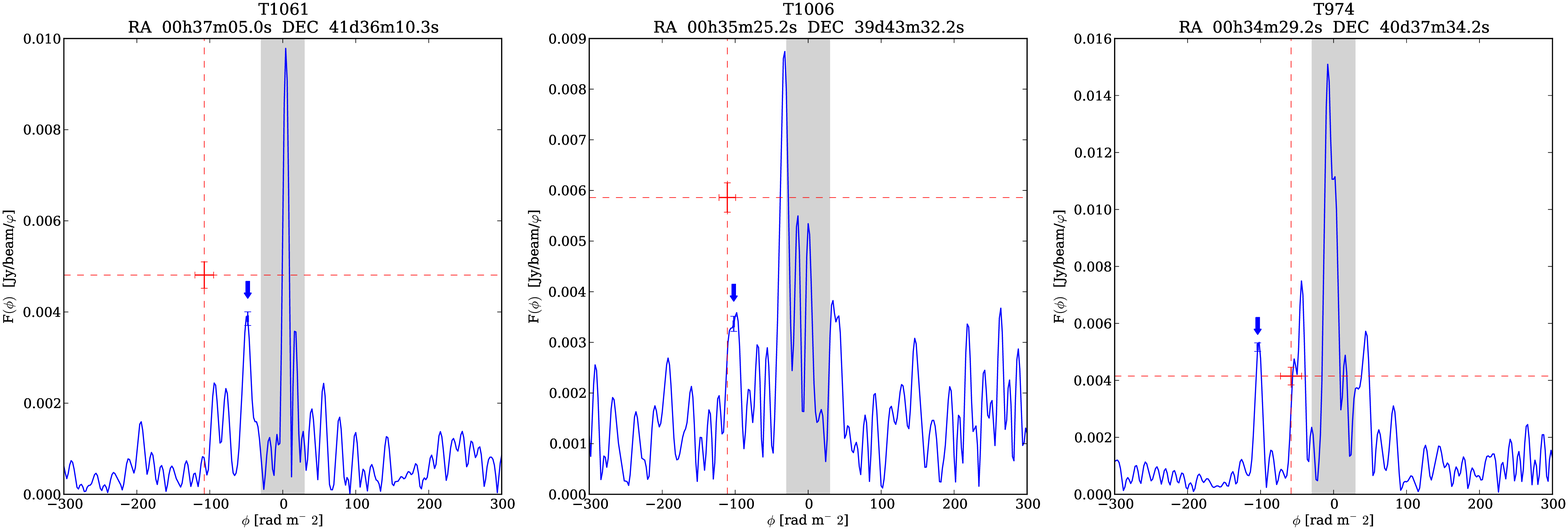}
\caption{Faraday Spectra of the detected sources (continued).}
\label{rm_spec6}
\end{center}
\end{figure}

\end{appendix}

\end{document}